\documentclass[10pt,a4paper,twocolumn]{article}
\usepackage{f1000_styles}

\definecolor{paleblue}{HTML}{7892A4}
\definecolor{darkblue}{HTML}{45596E}
\definecolor{midblue}{HTML}{4D637A}

\usepackage{graphicx,wrapfig}

\usepackage{enumitem}
\usepackage{hyperref}
\hypersetup{colorlinks,linktocpage,linkcolor={darkblue}, citecolor={darkblue}, urlcolor={darkblue}}

\usepackage{atlast}

\newcounter{naffil}
\usepackage{bm}

\newcommand*{\addaffil}[1]{
    \refstepcounter{naffil}\affil[\thenaffil]{#1}
}

\usepackage{rotating}

\begin{document}
\pagestyle{fancy}

\title{Atacama Large Aperture Submillimeter Telescope (AtLAST) science: Resolving the hot and ionized Universe through the Sunyaev-Zeldovich effect}

\author[1-4]{Luca Di Mascolo}
\author[5]{Yvette Perrott}
\author[6]{Tony Mroczkowski}
\author[7]{Srinivasan Raghunathan}
\author[8]{Stefano Andreon}
\author[9,10]{Stefano Ettori}
\author[11-13]{Aurora Simionescu}
\author[6,12]{Joshiwa van Marrewijk}
\author[14]{Claudia Cicone}
\author[15,16]{Minju Lee}
\author[17]{Dylan Nelson}
\author[18,19]{Laura Sommovigo}
\author[20]{Mark Booth}
\author[20]{Pamela Klaassen}
\author[6]{Paola Andreani}
\author[21]{Martin A.\ Cordiner}
\author[22,23]{Doug Johnstone}
\author[6]{Eelco van Kampen}
\author[24,25]{Daizhong Liu}
\author[26]{Thomas J.\ Maccarone}
\author[27,28]{Thomas W.\ Morris}
\author[29]{John Orlowski-Scherer}
\author[30,31]{Am\'elie Saintonge}
\author[32]{Matthew Smith}
\author[33]{Alexander E.\ Thelen}
\author[14,34]{Sven Wedemeyer}

\addaffil{Kapteyn Astronomical Institute, University of Groningen, Landleven 12, 9747 AD, Groningen, The Netherlands}
\addaffil{Laboratoire Lagrange, Université Côte d'Azur, Observatoire de la Côte d'Azur, CNRS, Blvd de l'Observatoire, CS 34229, 06304 Nice cedex 4, France}
\addaffil{Astronomy Unit, Department of Physics, University of Trieste, via Tiepolo 11, Trieste 34131, Italy} 
\addaffil{INAF -- Osservatorio Astronomico di Trieste, via Tiepolo 11, Trieste 34131, Italy}

\addaffil{Victoria University of Wellington, Wellington, New Zealand}

\addaffil{European Southern Observatory (ESO), Karl-Schwarzschild-Strasse 2, Garching 85748, Germany}

\addaffil{Center for AstroPhysical Surveys, National Center for Supercomputing Applications, Urbana, IL 61801, USA}

\addaffil{INAF -- Osservatorio Astronomico di Brera, via Brera 28, 20121, Milano, Italy}

\addaffil{INAF -- Osservatorio di Astrofisica e Scienza dello Spazio, via Piero Gobetti 93/3, 40129 Bologna, Italy}
\addaffil
{INFN -- Sezione di Bologna, viale Berti Pichat 6/2, 40127 Bologna, Italy}

\addaffil{SRON, Netherlands Institute for Space Research, Niels Bohrweg 4, NL-2333 CA Leiden, the Netherlands}
\addaffil{Leiden Observatory, Leiden University, P.O. Box 9513, 2300 RA Leiden, The Netherlands}
\addaffil{Kavli Institute for the Physics and Mathematics of the Universe, The University of Tokyo, Kashiwa, Chiba 277-8583, Japan}

\addaffil{Institute of Theoretical Astrophysics, University of Oslo, P.O. Box 1029, Blindern, 0315 Oslo, Norway}

\addaffil{Cosmic Dawn Center (DAWN), Denmark}
\addaffil{DTU-Space, Technical University of Denmark, Elektrovej 327, DK2800 Kgs. Lyngby, Denmark}

\addaffil{Universität Heidelberg, Zentrum für Astronomie, Institut für Theoretische Astrophysik, Albert-Ueberle-Str. 2, 69120 Heidelberg, Germany}

\addaffil{Center for Computational Astrophysics, Flatiron Institute, 162 5th Avenue, New York, NY 10010, USA}
\addaffil{Scuola Normale Superiore, Piazza dei Cavalieri 7, I-56126 Pisa, Italy}

\addaffil{UK Astronomy Technology Centre, Royal Observatory Edinburgh, Blackford Hill, Edinburgh EH9 3HJ, UK}

\addaffil{Astrochemistry Laboratory, Code 691, NASA Goddard Space Flight Center, Greenbelt, MD 20771, USA.}

\addaffil{NRC Herzberg Astronomy and Astrophysics, 5071 West Saanich Rd, Victoria, BC, V9E 2E7, Canada}
\addaffil{Department of Physics and Astronomy, University of Victoria, Victoria, BC, V8P 5C2, Canada}

\addaffil{Max-Planck-Institut f\"{u}r extraterrestrische Physik, Giessenbachstrasse 1 Garching, Bayern, D-85748, Germany}
\addaffil{Purple Mountain Observatory, Chinese Academy of Sciences, Nanjing 210023, China}

\addaffil{Department of Physics \& Astronomy, Texas Tech University, Box 41051, Lubbock TX, 79409-1051, USA }

\addaffil{Department of Physics, Yale University, New Haven, CT 06511, USA}
\addaffil{Brookhaven National Laboratory, Upton, NY 11973, USA}

\addaffil{Department of Physics and Astronomy, University of
Pennsylvania, 209 South 33rd Street, Philadelphia, PA, USA 19104}

\addaffil{Department of Physics and Astronomy, University College London, Gower Street, London WC1E 6BT, UK}
\addaffil{Max-Planck-Institut f\"{u}r Radioastronomie (MPIfR), Auf dem H\"{u}gel 69, D-53121 Bonn, Germany}

\addaffil{School of Physics \& Astronomy, Cardiff University, The Parade, Cardiff CF24 3AA, UK}

\addaffil{Division of Geological and Planetary Sciences, California Institute of Technology, Pasadena, CA 91125, USA.}

\addaffil{Rosseland Centre for Solar Physics,  University of Oslo, Postboks 1029 Blindern, N-0315 Oslo, Norway}

\maketitle
\thispagestyle{fancy}

\clearpage
\thispagestyle{fancy}

\begin{abstract} 
An omnipresent feature of the multi-phase ``cosmic web'' --- the large-scale filamentary backbone of the Universe --- is that warm/hot ($\gtrsim 10^{5}~\mathrm{K}$) ionized gas pervades it. This gas constitutes a relevant contribution to the overall universal matter budget across multiple scales, from the several tens of Mpc-scale intergalactic filaments, to the Mpc intracluster medium (ICM), all the way down to the circumgalactic medium (CGM) surrounding individual galaxies from $\sim 1$~kpc up to their respective virial radii ($\sim 100$~kpc). 
The study of the hot baryonic component of cosmic matter density represents a powerful means for constraining the intertwined evolution of galactic populations and large-scale cosmological structures, for tracing the matter assembly in the Universe and its thermal history. To this end, the Sunyaev-Zeldovich (SZ) effect provides the ideal observational tool for measurements out to the beginnings of structure formation. The SZ effect is caused by the scattering of the photons from the cosmic microwave background off the hot electrons embedded within cosmic structures, and provides a redshift-independent perspective on the thermal and kinematic properties of the warm/hot gas. Still, current and next-generation (sub)millimeter facilities have been providing only a partial view of the SZ Universe due to any combination of: limited angular resolution, spectral coverage, field of view, spatial dynamic range, sensitivity, or all of the above.
In this paper, we motivate the development of a wide-field, broad-band, multi-chroic continuum instrument for the Atacama Large Aperture Submillimeter Telescope (AtLAST) by identifying the scientific drivers that will deepen our understanding of the complex thermal evolution of cosmic structures. On a technical side, this will necessarily require efficient multi-wavelength mapping of the SZ signal with an unprecedented spatial dynamic range (from arcsecond to tens of arcminutes) and we employ detailed theoretical forecasts to determine the key instrumental constraints for achieving our goals.
\end{abstract}

\clearpage
\pagestyle{fancy}

\section*{Plain language summary}
The matter content of the Universe is organized along a large-scale filamentary ``cosmic web'' of galaxies, gas, and an unseen ``dark matter'' component. Most of the ordinary matter exists as a diffuse plasma, with temperatures of $10-100$ million degrees. The largest concentrations of such gas are found within gigantic galaxy clusters at the intersections of the cosmic web, weighing as much as a few quadrillion Suns. As such, their masses, number and distribution across cosmic time probe the evolution and composition of the Universe itself. Also, the thermodynamics of the warm/hot gas provides an archaeological record of those mechanisms that, over time, influence the large-scale structures --- from the energy deposition by supermassive black holes to the collisions of massive clusters.

The best option for studying the cosmic warm/hot gas, especially at the beginnings of their formation, is provided by the so-called Sunyaev-Zeldovich (SZ) effect --- a faint distortion of the Cosmic Microwave Background (CMB) observable at (sub)millimeter wavelengths. Using the SZ effect to study cosmic thermal history however requires technical advances not met by state-of-the-art (sub)millimeter telescopes. In fact, many key questions on the co-evolution of the warm/hot gas, the embedded galaxies, and the cosmic web remain unanswered.

With these motivations in mind, we discuss the development of the Atacama Large Aperture Submillimeter Telescope (AtLAST). Thanks to its unprecedented combination of a 50-meter aperture and wide $2^\circ$ field of view (4$\times$ wider than the full Moon), AtLAST will map vast sky areas and detect extremely faint signals across multiple wavelengths. Overall, AtLAST will push these studies beyond the legacy of the many (sub)millimeter facilities that, during the 50 years since the theoretical foundations of the SZ effect, have pioneered the exploration of the warm/hot Universe through the SZ effect.

\section{Introduction}\label{sec:introduction}
\subsection{Clusters and the evolution of the large-scale structure of the Universe}

Clusters of galaxies, groups, and massive galaxies trace the large scale structure of the Universe, and have therefore, since their discovery, served as probes of cosmology \citep{Kravtsov2012}. For example, clusters provided the first tentative hints of dark matter \citep{Zwicky1933} as well as early evidence that we live in a universe with a low matter density $\Omega_M \sim 0.2-0.3$ \citep{Bahcall1992, White1993}. While the large catalogs compiled by cluster and large scale structure surveys have offered the testbeds of, for example, the growth of structure and cosmic shear \citep{Kilbinger2015,Mandelbaum2018,Huterer2023}, these tests are limited by systematics originating primarily from astrophysical effects --- shocks \citep[e.g.,][]{Markevitch2007}, feedback \citep[e.g.,][]{HlavacekLarrondo2022}, non-thermal pressure, and the objects' dynamical and virialization states \citep[e.g.,][]{Sullivan2024}, to name a few --- as well as from contamination due to interlopers and sources within the systems that can bias our measurements and any resulting cosmologically relevant observable like the cluster mass, a key proxy of structure evolution (see e.g., \citealt{Kravtsov2012, Pratt2019}).

Meanwhile, the same sources that can contaminate measurements, primarily radio-loud active galactic nuclei (AGN) and star forming galaxies, or cause departures from thermal equilibrium, such as shocks, are also the main drivers of the physical and thermodynamic evolution of the intracluster medium (ICM). The ICM, in turn, is the large scale environment within which a large fraction of galaxies reside, so the feedback and interactions between the two are important to both cosmology as well as cluster and galaxy evolution. The study of galaxy clusters is thus complex and multifaceted. Yet, constraining in greater detail the multi-scale physical processes taking place within the most massive structures in the Universe will ultimately allow us to build a more complete understanding of the thermal history of our universe, how structure grew and evolved, or fundamentally how the Universe came to be the way it is. Ultimately, by understanding the nature of clusters and large scale structure, we will also be able to test the properties of the ubiquitous dark matter \citep{Clowe2006}, and to peer into the dark universe itself.

\begin{figure*}[!ht]
    \centering
    \includegraphics[clip,trim=0cm 15cm 0cm 10cm,width=0.8\textwidth]{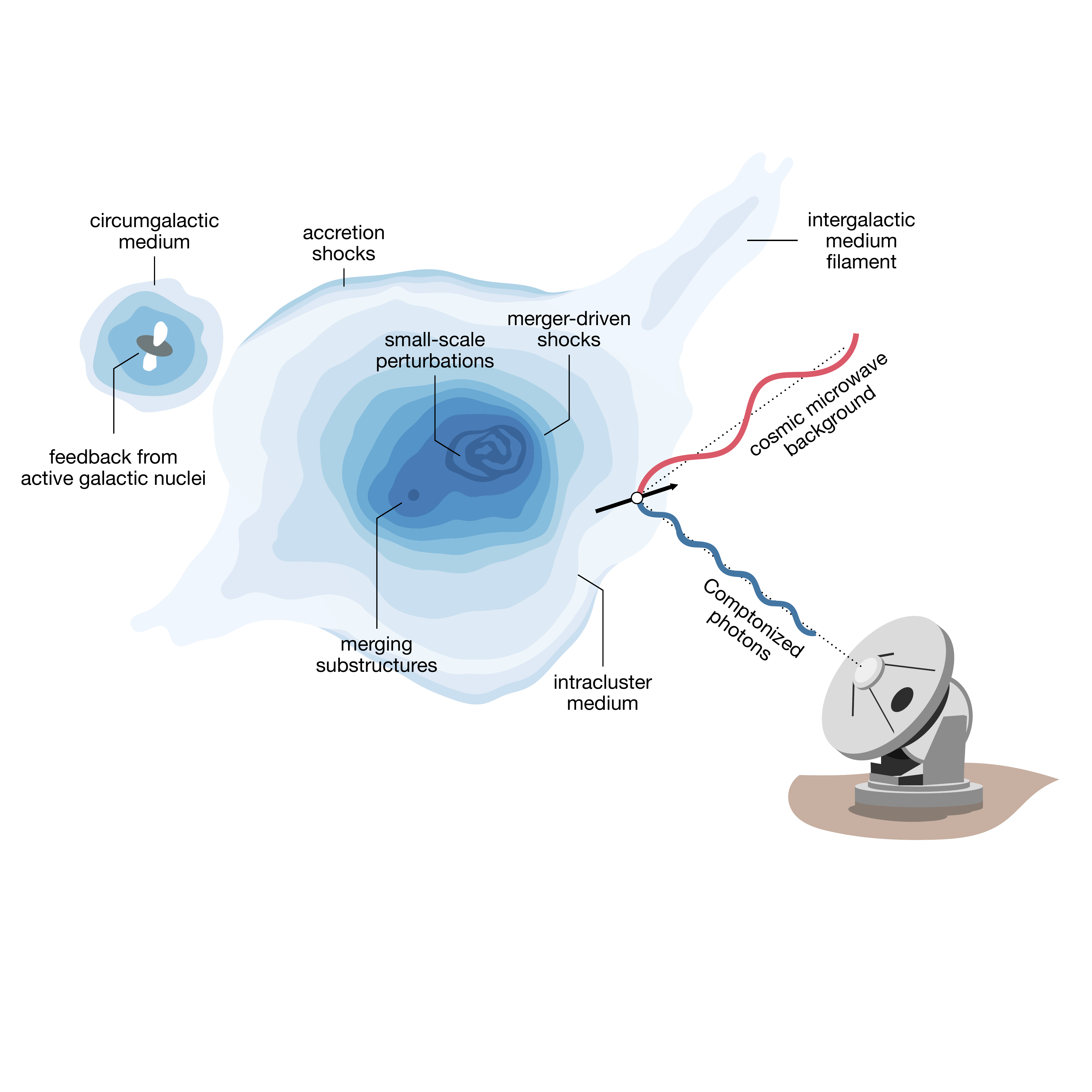}
    \caption{Expanded diagram highlighting some of the aspects of galaxy clusters and large-scale structures that will be studied through the Sunyaev-Zeldovich (SZ) effect using AtLAST. The SZ effect is caused by the interaction of photons from the cosmic microwave background (CMB) with reservoirs of energetic electrons within cosmic large-scale structures. Thanks to AtLAST's unparalleled capabilities, it will be possible to fully exploit multiple aspects of the SZ effect to characterize: the multi-scale properties of the intracluster medium (ICM; Sect.~\ref{sec:theory:profile}-\ref{sec:theory:ksz}) and the many dynamical processes likes mergers and large-scale accretion events shaping the observational properties of galaxy clusters; the elusive low-mass and high-redshift end of cosmic haloes (Sect.~\ref{sec:theory:elusive}-\ref{sec:theory:proto}); how active galactic nuclei (AGN) impact the thermal properties of the the circumgalactic medium (CGM) and of the ICM (Sect.~\ref{sec:theory:AGNfeedback}); how the warm/hot component of cosmic baryons is distributed within cluster outskirts and the large-scale filaments of intergalactic medium (IGM; Sect.~\ref{sec:theory:whim}). The figure is an adaptation of the SZ schematic in \citet{Mroczkowski2019}, which was based on that from L.\ van Speybroeck as adapted by J.\ E.\ Carlstrom.}
    \label{fig:sz_diagram}
\end{figure*}

\subsection{AtLAST, in brief}
With the ambitious goal of deeply improving our understanding of the formation and evolution of cosmic structures, we seek here to motivate deep, multi-band or multi-chroic high resolution and wide field observations with the Atacama Large Aperture Submillimeter Telescope \citep[\href{http://atlast-telescope.org/}{AtLAST};][]{Klaassen2020,Ramasawmy2022,Mroczkowski2023,Mroczkowski2025}. AtLAST is a proposed (sub)millimeter facility planned to be built on the Chajnantor Plateau. Taking advantage of the excellent atmospheric transmission high in the Atacama Desert, AtLAST will operate across the $30-950~\mathrm{GHz}$ spectral range. The planned 50-meter class aperture will allow observers to map the (sub)millimeter sky at high angular resolution, reaching $\approx 10~\mathrm{arcsec}$ at $150~\mathrm{GHz}$ and $\approx 1.5~\mathrm{arcsec}$ at $950~\mathrm{GHz}$. At the same time, AtLAST will feature an unparallel mapping speed, aiming at achieving an instantaneous field of view of $2~\mathrm{deg}$ diameter thanks to its novel optical design which will be coupled to next-generation large-format receivers, each expected to comprise up to $\sim10^6$ detector elements. With the goal of building a multi-purpose facility that could support the heterogeneous science applications of the (sub)millimeter community, AtLAST is planned to simultaneously host six different instruments, allowing for swift and efficient switching across them (see the AtLAST Memos \href{https://www.atlast.uio.no/memo-series/memo-public/memo3_instrument_mounting_options.pdf}{3} and \href{https://www.atlast.uio.no/memo-series/memo-public/instrumentationwgmemo4_29feb2024.pdf}{4} for a summary). Such ambitious requirements are not met by any present facilities or any of their future upgrades, and have served as key technical drivers for the ongoing design effort \citep{Klaassen2020}, starting with the preliminary optical design concept outlined in the AtLAST Memos \href{https://www.atlast.uio.no/memo-series/memo-public/basic-layout-options-v2.pdf}{1} and \href{https://www.atlast.uio.no/memo-series/memo-public/wobbler_for_atlast.pdf}{2} by R.\ Hills. 
We also note that the AtLAST aims to serve as a novel, sustainable facility, both from an energetic and social perspective \citep{Viole2023,Viole2024,ValenzuelaVenegas2024,Kiselev2024}.
The full design concept has recently been presented in a series of dedicated papers \citep{Gallardo2024a,Kiselev2024,Puddu2024,Reichert2024,Mroczkowski2025}, to which we refer the reader for more details.

The unprecedented combination of spectral coverage, angular resolution, and mapping speed will position AtLAST as a leading (sub)millimeter facility for addressing questions of cluster astrophysics as well as the contamination that could potentially plague cluster cosmology done at arcminute resolutions. At the same time, the observations discussed here are not simply to aid cosmological studies, but can probe interesting astrophysics and solve important questions about astrophysics in their own right --- with the unique potential of providing a link between galaxy evolution, large-scale structure, and cosmological studies.  Our primary tool here is the Sunyaev-Zeldovich effect, described below.

\section{The multi-faceted Sunyaev-Zeldovich effect}\label{sec:sz}
The Sunyaev-Zeldovich (SZ) effects \citep{Sunyaev1970, Sunyaev1972, Sunyaev1980} are caused by the up-scattering to higher energies of photons from the cosmic microwave background (CMB) by populations of free energetic electrons within cosmic structures. Depending on the specific velocity distribution of the scattering electrons, such an interaction imprints specific spatial and spectral variation in the surface brightness of the back-lighting CMB. As such, disentangling the relative contribution from each SZ component helps us exploit the different information they carry to study the physical properties of the ionized gas throughout the Universe. In this section we provide a brief introduction to the specific SZ contributions that we aim at probing with AtLAST that will be key for studying the varied astrophysical processes occurring in clusters and large-scale structures (we refer to Figure~\ref{fig:sz_diagram} for a cartoon depiction, and to the subsections of Sect.~\ref{sec:theory} for a detailed discussion in the AtLAST context). For more comprehensive reviews of the various aspects of the SZ effect, we further refer to, e.g., \citet{Birkinshaw1999, Carlstrom2002, Kitayama2014}, and \citet{Mroczkowski2019}.

\subsection{Thermal component}\label{sec:sz:tsz}
The thermal SZ effect \citep{Sunyaev1970, Sunyaev1972} was proposed theoretically as an alternative to X-ray measurements to probe the thermodynamics of the hot gas in galaxy clusters. It is produced by the scattering of CMB photons by a reservoir of hot electrons in thermal equilibrium resulting in a spectral distortion of the CMB. For a given direction $\bm{n}$ on the sky, the classical (non-relativistic) thermal SZ effect induces a variation in the CMB temperature $T_{\textsc{cmb}}$ with amplitude
\begin{equation}
    \textstyle
    \frac{\Delta T_{\textsc{cmb}}(\bm{n};x)}{T_{\textsc{cmb}}} = g(x)~y_{\mathrm{t\textsc{sz}}}(\bm{n}) = \left[x\coth{\left(\frac{x}{2}\right)}-4\right] y_{\mathrm{t\textsc{sz}}}(\bm{n}),
    \label{eq:ytsz_1}
\end{equation}
for a given dimensionless frequency $x=h\nu/k_{\textsc{b}}T_{\textsc{cmb}}$, with $h$ equal to the Planck constant, $\nu$ the photon frequency, and $k_{\textsc{b}}$ the Boltzmann constant. The second term in the equation denotes the thermal Compton-$y$ parameter (hereafter, $y_{\mathrm{t\textsc{sz}}}$), proportional to the integral along the line of sight $l$ of the electron pressure distribution $P_{\mathrm{e}}(\bm{n},l)$,
\begin{equation}
    \textstyle
    y_{\mathrm{t\textsc{sz}}}(\bm{n}) = \frac{\sigma_{\textsc{T}}}{m_{\mathrm{e}}c^2} \int P_{\bm{e}}(\bm{n},l) \mathrm{d}l.
    \label{eq:ytsz_2}
\end{equation}
Here, $\sigma_{\textsc{t}}$, $m_{\mathrm{e}}$, and $c$ denote the Thomson cross section, electron mass, and speed of light, respectively. In other terms, the thermal SZ effect provides a direct proxy for the (thermal) pressure due to the free electrons in the ICM and, as such, the optimal tool for gaining a direct calorimetric view of the gas thermal properties.

\subsection{Kinetic component}\label{sec:sz:ksz}
Shortly after the theoretical foundations of the thermal SZ effect, the kinetic SZ effect \citep{Sunyaev1980} was proposed as a way to measure gas momentum with respect to the CMB, our ultimate and most universal reference frame. Given a collection of electrons, characterized by a number density $n_{\mathrm{e}}(\bm{n},l)$, that are moving at a velocity with a component along the line of sight $\beta_{l}(\bm{n},l)$ (in units of speed of light), the CMB temperature variation is given by:
\begin{equation}
    \textstyle
    \frac{\Delta T_{\textsc{cmb}}(\bm{n})}{T_{\textsc{cmb}}} = - \sigma_{\textsc{t}} \int  n_{\mathrm{e}}(\bm{n},l) \beta_{l}(\bm{n},l) \mathrm{d}l = -y_{\mathrm{k\textsc{sz}}}(\bm{n}).
    \label{eq:ksz}
\end{equation}
For the sake of consistency with literature (e.g., \citealt{Ruan2013,Adam2017,Biffi2022,MonllorBerbegal2024}) and of facilitating the comparison with the thermal SZ effect, we introduce here a pseudo-Compton-$y$ parameter for the kinetic SZ effect $y_{\mathrm{k\textsc{sz}}}=\sigma_{\textsc{t}}\int n_{\mathrm{e}} \beta_{l}\mathrm{d}l$ in analogy to the thermal SZ $y_{\mathrm{t\textsc{sz}}}$. We note that, while $y_{\mathrm{k\textsc{sz}}}$ is not effectively equivalent to a Compton-$y$ parameter, it still provides a practical reference for the magnitude of the kinetic SZ signal when comparing it to any corresponding thermal SZ signal.

\subsection{Relativistic corrections and non-thermal component}\label{sec:sz:rsz}
\begin{figure*}
    \begin{center}
    \includegraphics[clip, trim=15mm 0mm 5mm 5mm, width=0.85\textwidth]{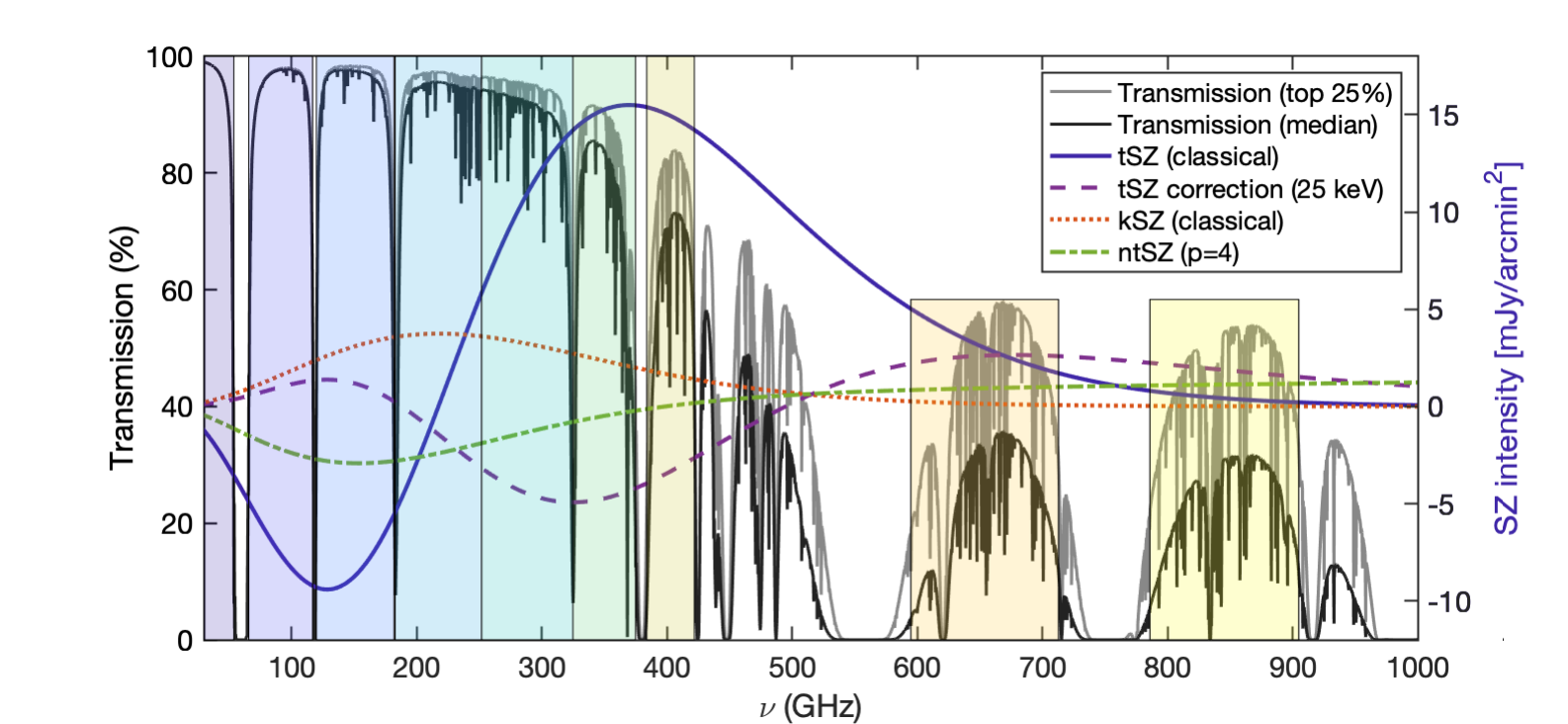}
    \vspace{-10pt}
    \end{center}
    \caption{Various SZ spectra versus transmission in the top quartile (lighter gray) and median (darker gray) atmospheric transmission conditions available at the Chajnantor Plateau ($\approx 5000$~meters above sea level). The left y-axis corresponds to transmission, and the right y-axis is appropriate for the thermal SZ (tSZ) intensity for a cluster with $y_{\mathrm{t\textsc{sz}}}=10^{-4}$. The kinetic SZ (kSZ) values assume a line of sight velocity component $v_{l} = -1000~\mathrm{km/s}$ (i.e.\ toward the observer, implying a net blueshift in the CMB toward the cluster) and an electron opacity $\tau_{\mathrm{e}} = 0.01$. The non-thermal SZ term (ntSZ) is computed assuming a population of mono-energetic relativistic electrons with normalized momentum $p=4$. In all cases, we used \texttt{SZpack} to solve for the SZ spectral distortions \citep{Chluba2012}, and the \texttt{am} code for atmospheric transmission \citep{Paine2019}.
    The optimal continuum bands of the proposed AtLAST SZ observations, reported in Table~\ref{tab:freq_sens_beam}, are shown as background shaded regions.}
    \label{fig:sz_spectrum}
\end{figure*}

The decade following the first prediction of the thermal and kinetic SZ effects saw developments in the theory regarding relativistic corrections to the thermal SZ and kinetic SZ effects, as well as anticipating more exotic SZ effects from non-thermal and ultra-relativistic electron populations \citep[e.g.,][]{Itoh1998,Nozawa1998,Ensslin2000,Colafrancesco2003,Chluba2012,Chluba2013}. In the single scattering approximation (see \citealt{Chluba2014a} and \citealt{Chluba2014b} for an extension to case of multiple scatterings), a static and isotropic distribution of electrons with optical depth $\tau_{\mathrm{e}}(\bm{n})=\sigma_{\textsc{t}}\int{n_{\mathrm{e}}(\bm{n},l)\mathrm{d}l}$ induces a distortion of the measured temperature of the underlying CMB with amplitude:
\begin{equation}
    \textstyle
    \frac{\Delta T_{\textsc{cmb}}(\bm{n};x)}{T_{\textsc{cmb}}} = \frac{4\sinh^2{x}}{x^4} [j(x)-j_0(x)] \tau_{\mathrm{e}},
    \label{eq:rtsz1}
\end{equation}
where $j_0(x) = x^3~(e^x-1)^{-2}$ is the normalized blackbody spectrum at a given dimensionless frequency $x$. The quantity $j(x)$ encompasses the full information regarding the momentum distribution of the scattering electrons and the relativistic treatment of the inverse Compton scattering process itself,
\begin{equation}
    \textstyle
    j(x) = \int_0^{\infty} j_0(x/t) \int_0^{\infty} f_{\mathrm{e}}(p) P(t;p)~\mathrm{d}p ~\mathrm{d}t.
    \label{eq:jx}
\end{equation}
Here, we introduced the redistribution function $P(t;p)$ describing the probability for a mono-energetic electron distribution to upscatter a photon to $t$ times its original frequency (see \citealt{Ensslin2000} for an analytical expression in the single scattering limit). The term $f_{\mathrm{e}}(p)$ instead denotes the electron momentum distribution, assuming an electron with speed $v=\beta c$ and a dimensionless electron momentum $p=\beta\gamma=\beta/\sqrt{1-\beta^2}$. Different electron populations with a specific spectrum $f_{\mathrm{e}}(p)$ would thus result in characteristic spectral distortion. For instance, in the case of an electron reservoir in thermal equilibrium at a temperature $T_{\mathrm{e}}$, $f_{\mathrm{e}}(p)$ follows a relativistic Maxwell-Boltzmann distribution, i.e., $f_{\mathrm{e}}(p;T_{\mathrm{e}})\propto \beta_{\mathrm{e}} p^2 \exp{(-\beta_{\mathrm{e}}\sqrt{1+p^2})}$ with $\beta_{\mathrm{e}}=m_{\mathrm{e}}c^2/k_{\textsc{b}}T_{\mathrm{e}}$. This implies that the spectral scaling $j(x)$ becomes directly dependent on the specific temperature of the scattering electrons and deviates from the non-relativistic thermal SZ case introduced in Sect.~\ref{sec:sz:tsz}. As an example, we report in Fig.~\ref{fig:sz_spectrum} the relativistic correction to the thermal SZ spectrum in the specific case of $T_{\mathrm{e}}=25~\mathrm{keV}$. The net result is an overall reduction of the amplitude of the SZ decrement and increment peaks, along with a shift of the crossover point (i.e., $\Delta T_{\textsc{cmb}}=0$) to higher frequencies. Lower and higher temperatures suppress and accentuate such effects, respectively, thus allowing one to use any deviation from the non-relativistic thermal SZ spectrum (Sect.~\ref{sec:sz:tsz}) as a direct observational probe of the electron temperature $T_{\mathrm{e}}$. In this regard, we note that, over the past few decades, several authors have worked on numerical integrations and asymptotic expansions of the scattering equations, with the specific aim of simplifying the estimation of the scattering integral (Eq.~\ref{eq:jx}) and, thus, of any temperature-dependent effect on the measured thermal SZ signal. This has resulted, for instance, in several effective formulations of the relativistic corrections to the thermal SZ effect in terms of electron temperature moments of varying order. We refer to \citet{Sazonov1998,Itoh1998,Challinor1998,Chluba2012,Lee2024SZ} for some relevant examples.

In addition to the above relativistic thermal corrections, non-thermal relativistic electrons --- from, e.g., aged radio plasmas responsible of the cluster-scale radio haloes \citep{vanWeeren2019} or AGN-driven relativistic outflows \citep{Fabian2012,Werner2019} --- are expected to introduce non-trivial contributions to the overall SZ effect (e.g., \citealt{Pfrommer2005,Colafrancesco2008,Prokhorov2010,Acharya2021,Muralidhara2024}). It is possible to gain a simple understanding of the expected SZ signature by considering a population of mono-energetic electrons $f_{\mathrm{e}}(p) = \delta(p-p_0)$. This simplifies $j(x)$ only to the integral over the blackbody component $j_0(x)$. An example derived under the assumption of $p=4$ is shown in Fig.~\ref{fig:sz_spectrum}, showing how the non-thermal SZ effect exhibits a different spectral shape than the thermal SZ effect, making them ideally separable via a spectral analysis.

Finally, we note that the discussion above is only strictly valid under the assumption of an isotropic CMB and scattering fields, and negligible systemic velocities. The inclusion of anisotropies in any of these components would introduce additional contributions to the total SZ spectral distortion, whose amplitudes can however be more than two orders of magnitude smaller than dominant thermal, kinetic, and relativistic SZ signals. These effects will be explored in future works and, for the moment, we refer the reader to, e.g., \citet{Mroczkowski2019}, \citet{Lee2024SZ}, and references therein for more details.

\subsection{Observational studies}
Despite the great theoretical effort, observations of the SZ effects took longer to come to the fore, beginning with pioneering measurements such as \citet{Birkinshaw1984} and culminating more recently in several thousand measurements or detections from low-resolution (1-10\arcmin) SZ surveys \citep[e.g.,][]{Planck2016,Bleem2020,Huang2020,Hilton2021,Bleem2023}. In the past decades, dedicated observations at high angular (subarcminute) resolution have started come to the fore, providing a detailed and direct perspective on the complex pressure structure of galaxy clusters \citep{Mroczkowski2019} --- ranging from the study of resolved pressure profiles \citep{Halverson2009,Adam2014,Ruppin2018,DiMascolo2019,Romero2020,Kitayama2020,Kitayama2023}, to the characterization of mergers  \citep{Mason2010,Plagge2013,Sayers2013,Kitayama2016,Basu2016,Adam2017,DiMascolo2019b,DiMascolo2021} and feedback-driven \citep{Abdulla2019,OrlowskiScherer2022} substructures, ICM turbulence \citep{Khatri2016,Romero2023,Romero2024b,Adam2025} and of systems in the earliest phases of cluster evolution (e.g., \citealt{Gobat2019,Andreon2023,DiMascolo2023,vanMarrewijk2023}).

The proposed millimeter/submillimeter facility, AtLAST, presents novel, unique capabilities that will revolutionize both deep targeted observations aiming for detailed astrophysical studies, as well as wide-field surveys aiming to push SZ observations to much lower mass limits and higher redshifts. Since the epoch of reionization, the majority of baryons have been making their way up to high enough temperatures ($>10^{5}~\mathrm{K}$) that their emission is nearly completely undetectable at optical and near-infrared wavelengths, where the majority of telescopes operate. Such a hot phase is an omnipresent feature of the multi-phase cosmic web, representing a relevant contribution to the volume-filling baryonic matter budget on multiple scales --- from Mpc-scale filaments of intergalactic medium (IGM), to the intracluster medium (ICM), and down to the circumgalactic medium (CGM) surrounding individual galaxies up to their virial radius (up to few 100s of kpc). Through the SZ effect, the millimeter/submillimeter wavelength regime offers a view of this important component of galaxies and their surrounding environments (clusters, groups, filaments) --- components that are largely invisible to all but X-ray and SZ instruments.

\section{Proposed science goals}\label{sec:theory}
Here we provide a summary of the main applications in the context of SZ studies enabled by AtLAST that will allow us to develop a more profound and complete understanding of the thermal history of the Universe, ultimately transforming our understanding of the numerous processes involved in structure formation, evolution, feedback, and the quenching of star formation in overdense environments. In this regard, we refer to \citet{Lee2024} and \citet{vanKampen2024} for companion AtLAST case studies focused on emission line probes of the cold circumgalactic medium (CGM) of galaxies and on providing a comprehensive survey of high-$z$ galaxies and protoclusters, respectively. Common to all the specific science cases discussed below is the need for a wide field, high angular resolution facility able to optimally probe the full SZ spectrum (Figure~\ref{fig:sz_spectrum}). More details about the optimal spectral setup, with a discussion on the optimal bands and corresponding noise performance are provided in Table~\ref{tab:freq_sens_beam}. We further refer to Sect.~\ref{sec:technical} below for a more extended discussion of the technical requirements for the proposed science goals.

\subsection{Thermodynamic properties of the ICM: radial profiles and small-scale perturbations}\label{sec:theory:profile}
The morphological and thermodynamic properties of the ICM represent key records of the many physical processes shaping the evolution of galaxy clusters and groups. Non-gravitational processes --- e.g., cooling, AGN feedback, different dynamical states and accretion modes \citep{Battaglia2012,Ghirardini2019} --- are expected to leave their imprint on the pressure distribution of the ICM in the form of deviations from the radial models derived under universal and self-similar assumptions for structure formation (see, e.g., \citealt{Nagai2007,Arnaud2010,Sayers2023}). On cluster scales, shock fronts induced by cluster mergers as well as cosmological accretion deposit their kinetic energy into the ICM, contributing to its overall thermalization \citep{Markevitch2007,Ha2018}.
On smaller scales, turbulent motion \citep{Schuecker2004,Khatri2016,Romero2023,Romero2024b,Adam2025} can induce significant non-thermal contributions to the ICM pressure support, in turn hampering the validity of the hydrostatic equilibrium assumption.
We thus need robust constraints on the level of turbulence affecting the energy budget of the ICM along with an independent census of the ``hydrostatic mass bias'' --- the systematic discrepancy between the true mass of a galaxy cluster and the value estimated from ICM-related proxies under the assumption of hydrostatic equilibrium (e.g., \citealt{Biffi2016}) --- via a combination of fluctuations and resolved hydrostatic mass information. This will be crucial for inferring corrections to the hydrostatic mass due to the non-thermalized gas (see, e.g., \citealt{Angelinelli2020,Ettori2022}) and therefore strengthening the role of thermodynamic quantities for cosmological purposes \citep{Pratt2019}. 

\begin{figure*}[!ht]
    \centering
    \includegraphics[width=0.98\textwidth]{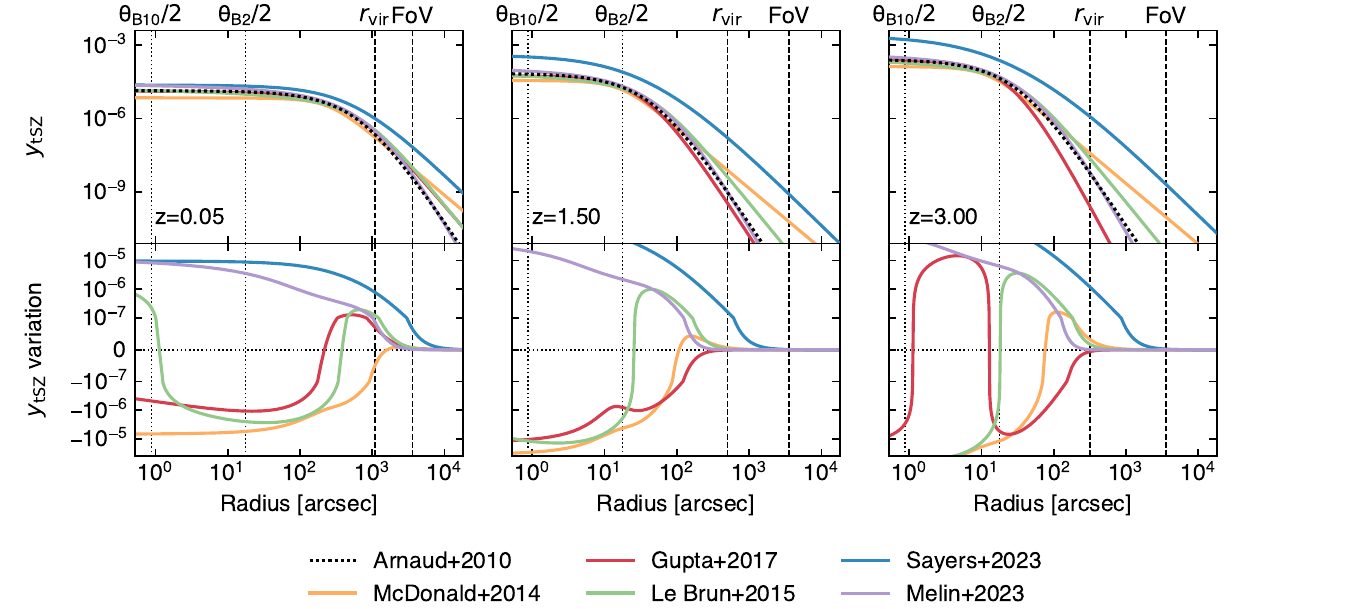}
    \caption{Observed thermal Compton $y_{\mathrm{t\textsc{sz}}}$ profiles when considering different fiducial models for the ICM pressure distribution (top panels; \citealt{Arnaud2010,McDonald2014,LeBrun2015,Gupta2017,Sayers2023,Melin2023}) and respective variations (bottom panels) with respect to the universal pressure profile from \citet{Arnaud2010}, commonly adopted as reference model for the inference of cluster masses. In this plot, we consider a cluster with fixed mass ($M_{\mathrm{500}}=10^{14}~\mathrm{M_{\odot}}$) for an arbitrary set of redshifts ($z=\{0.05,1.50,3.00\}$). The different profiles are computed on heterogeneous samples in terms of mass and redshift ranges, and thus encode different biases associated to the intrinsic scatter of pressure profiles, deviations from self-similar evolution and hydrostatic equilibrium. Thanks to AtLAST's sensitivity to Compton $y_{\mathrm{t\textsc{sz}}}$ levels $\lesssim10^{-7}$, it will be possible to characterize such effects, while providing a model for the evolution of ICM pressure across cosmic history. We note that the observed redshift evolution of the profiles is not only due to astrophysical effects or observational biases, but is also due to the assumption a fixed value for the cluster's $M_{500}$. Thus the redshift-dependent variation in amplitude and extent of the SZ profiles reflects the critical density and angular diameter distance for the redshift under consideration. As reference, we report as dashed vertical lines the virial radius of the model clusters and the instantaneous field of view expected for AtLAST (see Sect.~\ref{sec:technical}). We further denote as dotted vertical lines the largest and smallest angular resolution $\theta_{i}$ achievable with AtLAST, respectively obtained in the proposed Band 2 and Band 10 (see Sect.~\ref{sec:test:bands} and Table~\ref{tab:freq_sens_beam} below).
    }
    \label{fig:szprof}
\end{figure*}

As discussed in Sect.~\ref{sec:sz:tsz}, the thermal SZ effect provides an optimal probe of the pressure structure and thermal energy of the ICM. In fact, observational models for a statistically relevant sample of clusters are currently limited to the indirect determination of resolved pressure models for clusters up to $z \lesssim 1$ \citep{Arnaud2010,McDonald2014,Sayers2023}. Direct constraints of the properties of the ICM within protoclusters and clusters early in their formation have been obtained for only a handful of extreme systems ($z>1$; \citealt{Tozzi2015,Brodwin2016,Gobat2019,DiMascolo2023,Andreon2021,Andreon2023,vanMarrewijk2023}) or limited samples (e.g., \citealt{Ghirardini2021}). Despite the significant time investment with the Atacama Large Millimeter/Submillimeter Array (ALMA; \citealt{Wootten2009}) or the 100-meter Green Bank Telescope (GBT; \citealt{White2022}), these observations only allow one to perform a characterization of the physical and thermodynamic state of these early systems for a few select systems. Still, these have generally required the combination with ancillary X-ray observations, due to observational limitations including poor signal-to-noise ratio (S/N), the data being limited to fewer than 5 bands, and interferometric filtering effects 
(we note that ALMA is limited to the recovery of $\lesssim 1$ arcminute scales at $\nu > 100~\mathrm{GHz}$).

In order to gain a radially resolved view of pressure profiles and of their small-scale perturbations for a large variety of clusters (in terms of dynamical state, mass, and redshift), it is key to have simultaneous access to enhanced sensitivity, high angular resolution, and wide spectral coverage across the millimeter/submillimeter spectrum. These observations are important, as from hydrodynamical simulations the pressure distribution of high-$z$ galaxy clusters are predicted to diverge from the universal pressure models \citep{Battaglia2012,Gupta2017}, leading to a systematic offset between the mass-to-SZ observable scaling relation for high-$z$ haloes with respect to local ones \citep{Yu2015}. Constraining such deviations is crucial as they carry fundamental information on the complex interplay between all those multi-scale processes --- e.g., merger and accretion events, AGN and stellar feedback, turbulent motion --- at epochs ($z>1$) when their impact from galactic to cluster scales are expected to be the strongest. In Figure~\ref{fig:szprof}, we provide a comparison between a set of fiducial thermal SZ models derived either from cluster samples covering different mass and redshift ranges --- and, thus, encompassing varied levels of bias due to the intrinsic scatter of the pressure profiles, deviations from self-similar evolution and hydrostatic equilibrium --- or from heterogeneous simulations encoding different feedback prescriptions. Each model exhibits specific and characteristic radial behaviors, differing the others at levels that will be easily distinguishable by AtLAST.

At the same time, tracing the pressure profiles out to the cluster outskirts will be key to pinpoint and characterize virial and accretion shocks \citep{Hurier2019,Anbajagane2022,Anbajagane2023}, whose existence is a fundamental prediction of the current paradigm of large-scale structure formation \citep[e.g.,][]{Ryu2003,Zhang2021}. In particular, the location and properties of their SZ features can be exploited to study the mass assembly of galaxy clusters and to place direct constraints on their mass accretion rate (a quantity otherwise difficult to infer observationally; see, e.g., \citealt{Molnar2009,Lau2015,Baxter2021,Towler2023,Baxter2024}). Further, pressure perturbations due to the turbulent motion within the ICM have been measured to result in fluctuations of the Compton $y_{\mathrm{t\textsc{sz}}}$ signal with fractional amplitude $\lesssim10^{-1}$ compared to the underlying bulk SZ signal \citep{Khatri2016,Romero2023,Romero2024b,Adam2025}. The enhanced sensitivity and calibration stability that will be achieved by AtLAST will allow it to easily probe this level of fluctuations, providing important albeit indirect information on the level of non-thermal pressure support in the ICM (we refer to \citealt{Romero2024} for an extensive discussion). More in general, it is only with the unique technical prospects offered by AtLAST that we will be able to probe to thermal SZ signal down to the levels Compton $y_{\mathrm{tsz}} \approx 10^{-7}$  (Sect.~\ref{sec:test:det}) required to probe the full extent of the ICM pressure distribution (Figure~\ref{fig:szprof}), unparalleled by any of the current or forthcoming submillimeter facilities. To provide a simple visualization of AtLAST's expected capabilities, we show in left panel of Figure~\ref{fig:tsz-ksz} the thermal SZ signal for a simulated massive cluster extracted from the TNG-Cluster suite \citet{Nelson2024}. The Compton $y$ sensitivity that will be targeted by AtLAST will provide the means not only to probe the virial region of galaxy clusters, but also to access in a systematic way the fainter structures populating cluster outskirts. It is also worth noting how the different dependence of the thermal SZ effect and X-ray emission on the electron density --- linear and squared, respectively --- causes the thermal SZ effect to exhibit a dynamic range between the core and outskirt signals that is significantly smaller than the corresponding X-ray emission. It is precisely this property that makes the SZ effect particularly effective in the study of low-surface brightness features.

\subsection{Measuring the ICM temperature via relativistic SZ effect}\label{sec:theory:rtSZ}
As detailed in Sect.~\ref{sec:sz:rsz}, accounting for the relativistic velocities of hot electrons introduces a temperature-dependent distortion of the SZ spectral model (Figure~\ref{fig:sz_spectrum}). The resulting relativistic SZ effect thus offers a valuable (yet largely unexplored) opportunity to directly measure the temperature of ICM electrons. This represents a key ingredient for enhancing our physical models of galaxy clusters and improving their utility as cosmological probes via more accurate tuning of mass calibrations and scaling relations (e.g., \citealt{Lee2020,Remazeilles2020,Perrott2024}). At the same time, having simultaneous access to the full ICM thermodynamics (via temperature $T_{\mathrm{e}}$, as well as pressure $P_{\mathrm{e}}$ and density $n_{\mathrm{e}}$ measurements via the combination of the relativistic and purely thermal SZ effects) for a large sample of clusters across a wide range of masses and redshift offers the key chance of building a temporal census of the ICM entropy distribution ($\propto T_{\mathrm{e}}~ n^{-2/3}_{\mathrm{e}}$, or $\propto T^{5/3}_{\mathrm{e}} P^{-2/3}_{\mathrm{e}}$ when considering thermodynamic quantities directly probed by the SZ effect; \citealt{Voit2005b}). The many processes affecting cluster evolution --- e.g., AGN and stellar feedback, injection of kinetic energy due to merger activity --- are observed to modify the entropy profiles throughout the cluster volumes \citep[e.g.,][]{Pratt2010,Walker2012,Ghirardini2017}, compared to a baseline model that includes only the non-radiative sedimentation of low-entropy gas driven by gravity \citep{Tozzi2001,Voit2005}. As such, the spatially resolved study of the ICM entropy distribution provides a fundamental proxy of the thermal evolution of cosmic structures as well as the specific dynamical state of galaxy clusters.

Currently, estimates of the relativistic corrections to the thermal SZ effect are limited to a few pioneering studies targeting individual systems \citep{Hansen2002,Prokhorov2012} or focusing on stacking analyses \citep[e.g.,][]{Hurier2016,Erler2018,Remazeilles2025}. 
Still, even in the case of individual clusters with extremely rich observational spectral coverage (see, e.g., \citealt{Zemcov2012} and \citealt{Butler2022}, focusing on the well-known cluster RX~J1347.5-1154), SZ-based inferences of the ICM temperature have commonly resulted in constraints with limited significance.
Higher angular resolutions, such as those offered by AtLAST, will be an asset for constraining SZ temperatures. First, the higher angular resolution allows spatially-distinct foregrounds such as radio sources, dusty galaxies and the Galactic dust foreground to be accurately modelled and removed. Second, the extraction of resolved pressure and temperature profiles provides the unique opportunity of performing the physical modeling of the ICM relying solely on the SZ effect. 

Though access to X-ray-independent temperature constraints will likely be achievable only for a subset of massive haloes (see Sect.~\ref{sec:test:rtsz} below for details), such an SZ-driven approach will still carry a key advantage: they are not biased in the same way as measurements relying on X-ray spectroscopy. In contrast, electron temperatures measured using X-ray data are weighted roughly by the X-ray emission, which scales as density-squared (see e.g., \citealt{Mazzotta2004}) and is therefore subject to biases due to clumping and compression (e.g., \citealt{Simionescu2011}). Further, observations can become prohibitive at large cluster radii, due to the low X-ray emissivity, and at high redshift, due to cosmological dimming. It is worth noting that \citet{Churazov2015} showed that the self-similar evolution of galaxy clusters would introduce a near independence of redshift of the X-ray luminosity at fixed cluster mass --- when this is defined as the mass enclosed in the radius within which the average matter density equals some fiducial cosmic overdensity value (e.g., $500\times \rho_{\rm crit}$).  Nevertheless, we note that these considerations are valid only under the assumption that the local mass-observable scaling relations are applicable at high redshift. At the same time, both the resolved SZ signal and the respective cluster-integrated flux would still be $(1+z)^{3/2}$ larger than the X-ray emission from the same system at a given redshift $z$. On the other hand, since the SZ effect is characterized by a surface brightness that is inherently independent of redshift, ICM temperature constraints can in principle be derived without specific limits on the distance of the target systems.

Further, the temperatures inferred using data on the same clusters but taken using different X-ray observatories may suffer large systematic variations due to inherent calibration differences \citep{Schellenberger2015,Migkas2024}. In contrast, the SZ temperature estimate is pressure-weighted and is therefore predicted to be less biased by emission while being easier to constrain at large cluster radius due to the linear (instead of squared) dependence on density. And even in the case of low-mass (i.e., low-SZ surface brightness; see also Sect.~\ref{sec:theory:elusive}) clusters for which it will not be possible to extract resolved SZ-based temperature information, the availability of deep, high angular resolution SZ observations for a large sample of systems will still allow for matching resolution with X-ray observations and to extract resolved full thermodynamic properties of the ICM (see Sect.~\ref{sec:theory:profile} and Sect.~\ref{sec:synergy:xray}). 

An exploratory study of AtLAST's expected capabilities to measure temperature via the relativistic SZ effect is presented in Sect.~\ref{sec:test:rtsz}. We refer to this for more details on the impact of the specific spectral setup on the reconstruction of the relativistic SZ effect and on the technical requirements for extending such measurements over broad ranges of cluster masses and redshifts.

\subsection{Kinematic perspective on large scale structures}\label{sec:theory:ksz}
The kinetic component of the SZ effect represents a valuable tool for revealing the peculiar motion of cosmic structures (Sect.~\ref{sec:sz:ksz}). Nevertheless, its properties -- namely its shape, the fact that the kinetic SZ signal is generally weaker than the thermal SZ effect (Figure~\ref{fig:sz_spectrum}), and that it traces the integrated line of sight momentum -- make the kinetic SZ effect somewhat elusive to measure and interpret. Further, the kinetic SZ spectral signature is consistent with a Doppler shift of the CMB photons, making it spectrally indistinguishable from small-scale primordial CMB anisotropies. 

\begin{figure*}[!th]
    \centering
    \includegraphics[clip,trim=0 15pt 0 0,width=0.98\textwidth]{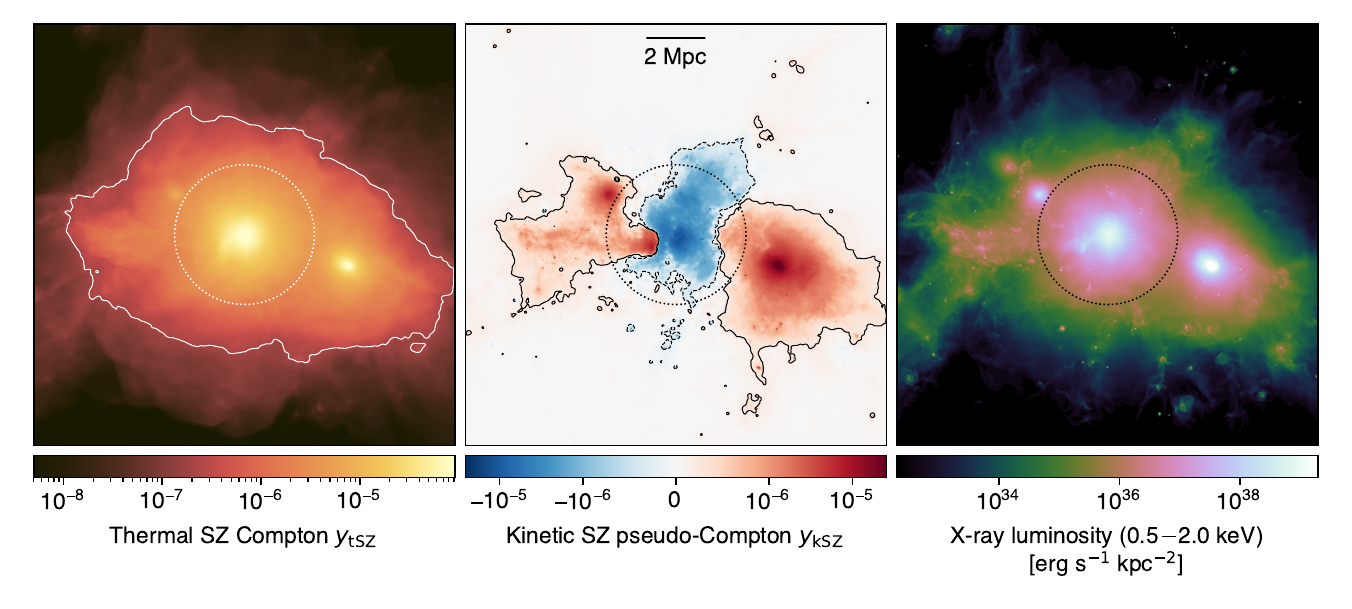}
    \caption{Thermal (left) and kinetic (center) SZ effects, and X-ray luminosity (right) from a simulated massive cluster undergoing a major merger ($M_{200}\simeq1.5\times10^{15}~\mathrm{M_{\odot}}$, $z=0$) extracted from the TNG-Cluster simulation (\citealt{Nelson2024}). The contours in both panels trace levels of $2\times10^{-7}$ both in $y_{\mathrm{t\textsc{sz}}}$ and $y_{\mathrm{k\textsc{sz}}}$, roughly corresponding to the reference SZ depth for a deep AtLAST survey (Sect.~\ref{sec:test:det}; see also Sect.~\ref{sec:introduction} for a definition of the thermal Compton $y_{\mathrm{t\textsc{sz}}}$ and kinetic pseudo-Compton $y_{\mathrm{k\textsc{sz}}}$ parameters). As a reference, we mark with a circle the virial radius of the galaxy cluster. This implies that AtLAST will be able to efficiently trace the SZ signal out to the low-density outskirts of clusters. At the same time, it will be possible to directly probe the velocity structure of individual galaxy clusters and their substructures.}
    \label{fig:tsz-ksz}
\end{figure*}

Past targeted kinetic SZ studies \citep[e.g.,][]{Mroczkowski2012,Sayers2013,Adam2017,Sayers2019,Silich2024a,Silich2024b} have already reported direct measurements of the kinetic SZ signal due to the large-scale gas flows associated with merger events. 
All of these works focused on individual, relatively extreme clusters (either in terms of overall mass, dynamical state, or orientation of the merger direction with respect to the line-of-sight). The broad spectral coverage and the expected sensitivity of AtLAST, in combination with its capability of probing a high dynamic range of angular scales, will instead allow for systematically including the kinetic SZ information in the reconstruction of the thermodynamic characterization of large statistical samples of galaxy clusters and groups. A simple illustration of AtLAST's view of the kinetic SZ effect from a massive system is provided in Figure~\ref{fig:tsz-ksz}, showing how the expected sensitivity will enable constraints on the bulk motion of individual cluster substructures.

\subsubsection{Cosmological applications}
Statistical measurements of the kinetic SZ effect in disturbed and merging systems represent a crucial ingredient for cosmological studies via direct measurements of the amplitude and the growth rate of cosmological density perturbations (e.g., \citealt{Bhattacharya2007,Soergel2018}), and have come to the fore in recent years by achieving ever increasing detection significances \citep[e.g.,][]{Amodeo2021, Calafut2021, Kusiak2021, Schaan2021, Vavagiakis2021, Hadzhiyska2024}, often by introducing novel methodologies \citep[e.g.,][]{McCarthy2024, Patki2024}.
Such statistical kSZ studies can also be used to distinguish $\Lambda$CDM from alternative cosmologies with modified gravitational forces \citep{Kosowsky2009,Mueller2015,Bianchini2016}.

We note that, while the small scales corresponding to spherical multipoles $\ell\approx7\,000$ are expected to be dominated by the kinetic SZ effect \citep{Smith2016}, the contamination from the CIB will limit the potential of existing and forthcoming CMB experiments to probe beyond  $\ell \gtrsim 5\,000$ \citep{Raghunathan2023}. AtLAST, on the other hand, when outfitted with the low-noise multi-band direct detection instrumentation we advocate for here, will be able to probe much smaller scales than dedicated CMB experiments improving the sensitivity to the kinetic SZ effect on small-scales (see the discussion in Sect.~\ref{sec:test:det:wide} below). For example, AtLAST will be able to make a robust measurement of the kinetic SZ power spectrum with $\mathrm{S/N}$ of a few hundreds, a least $2-3\times$ higher than what is expected from CMB-S4 \citep{Raghunathan2023}. Since the kinetic SZ signal contains contributions from the relativistic electrons within expanding ionizing bubbles during the epoch of reionization (EoR), the so-called ``patchy kinetic SZ'' signal (e.g., \citealt{Knox1998,McQuinn2005,Battaglia2013,Reichardt2016}), AtLAST's improved sensitivity to the kinetic SZ power spectrum on small scales will be crucial to constrain EoR \citep{Gorce2020, Reichardt2021, Gorce2022, Raghunathan2023}. 
In particular, AtLAST, using the full shape information of the kinetic SZ power spectrum, will be able to distinguish between different models of the patchy reionization process \citep{Jain2024}.

One of the limitations of using the kinetic SZ power spectrum is the challenge in disentangling the two main contributions to the large-scale kinetic SZ signal: the high-redshift ($z \gtrsim 6$) patchy reionization signal and the low-redshift ($z \lesssim 3$) component due to bulk peculiar motion of ionized gas in the local Universe. 
These two signals are expected to have roughly similar shape \citep{Shaw2012, Battaglia2013} and the current surveys do not have the sensitivity to distinguish between these two components, given the marginal detection significances of the kinetic SZ power spectrum from current surveys, which is further complicated by the foreground modeling \citep{Reichardt2021, Gorce2022}. 
AtLAST, with the high S/N detection of the kinetic SZ power spectrum and cross-correlation with galaxy surveys, will also be able to help in distinguishing between the two sources of the kinetic SZ signal.

\citet{Smith2016} proposed a new observable to mitigate the degeneracy between the two kinetic SZ sources for EoR studies. This method uses the trispectrum (4-point information) of the kinetic SZ signal which is expected to be dominated by the patchy reionization kinetic SZ effect \citep{Smith2016, Ferraro2018}. 
The technique was recently used by the South Pole Telescope (SPT) to set upper limits on the duration of EoR $\Delta_{z_{\rm re}} < 4.5$ (95 \% confidence level) with a non-detection of the kinetic SZ trispectrum signal \citep{Raghunathan2024}.
The main challenge for kinetic SZ trispectrum analysis with SPT \citep{Raghunathan2024} was the foreground mitigation and the multi-band configuration from AtLAST will greatly aid in improving this avenue.
The kinetic SZ power spectrum and trispectrum information from AtLAST when combined with other future probes like the large-scale reionization bump and the 21-cm observations, will further improve the EoR constraints significantly compared to what can be achieved by any of these probes individually.

\subsubsection{Resolved measurements}
The possibility of mapping the kinetic SZ effect at high angular resolution will open novel realms in the broader context of cluster studies. The resolved perspective will in fact be essential for correlating the kinetic information with the thermodynamic constraints based on the thermal SZ component and mitigate the spectral degeneracy of the kinetic SZ effect and the underlying primordial CMB anisotropies. On small scales, the velocity structure of the intracluster medium will introduce local (potentially non-gaussian) deviations in the CMB radiation field observed in the direction of a galaxy cluster, allowing for their separation via advanced summary statistical estimators and statistical component separation  (e.g., \citealt{Auclair2024,Blancard2024}).

Correlating the velocity structure with information from facilities at other wavelengths on the baryonic and dark matter content of merging systems will represent a preferential probe of the collisional nature of dark matter \citep{Silich2024a}. In the case of relatively relaxed systems (i.e., with velocity fields not manifesting complex morphologies), the joint analysis of the thermal and kinematic SZ effects would naturally complement the inference of the ICM pressure and temperature distributions with information on the bulk peculiar velocity of galaxy clusters and tighter constraints on the ICM density \citep{Mroczkowski2019}.

The detailed spatial mapping of the kinetic SZ effect could also be used to characterize turbulent motions and to identify their driving dissipation scales which are relevant for feedback mechanisms. This can be done, in particular, by computing the velocity structure function (VSF), defined as the average absolute value of the line of sight velocity differences as a function of projected scale separation. The VSF is an effective way of characterizing turbulent motions and identifying their driving and dissipation scales (see, e.g., \citealt{Li2020,Ganguly2023,Gatuzz2023,Ayromlou2023}). Determining the driving scale of turbulence would constrain the relative importance of gas motions driven by AGN feedback on small scales and mergers on large scales, while the dissipation scale is sensitive to the microphysics of the ICM, such as its effective viscosity \citep{Zhuravleva2019}. In general, constraints on the small-scale properties of the velocity field associated with turbulent motion \citep{Sunyaev2003,Nagai2003}, coherent rotation of gas within their host dark matter haloes \citep{Cooray2002,Baldi2018,Baxter2019,Altamura2023,Bartalesi2023}, or merger-induced perturbations \citep{Biffi2022} can complement the reconstruction of ICM thermodynamic fluctuations \citep{Khatri2016,Romero2023,Romero2024b,Adam2025} and the potential mitigation of biases due to non-thermal pressure support (e.g., \citealt{Shi2016,Angelinelli2020,Ansarifard2020,Ettori2022,Sullivan2024}) discussed in Sect.~\ref{sec:theory:profile}. Perturbations in the kinetic SZ distribution will result in small-scale kinetic SZ fluctuations more than an order of magnitude smaller than the corresponding thermal SZ component \citep{Sunyaev2003,Mroczkowski2019,Biffi2022} even for massive systems. The clear requirement of extremely demanding observations (along with the difficulty in spectrally disentangling the kinetic SZ effect from the underlying CMB signal; \citealt{Mroczkowski2019}) have so far limited the possibility of directly measuring any small-scale kinetic SZ feature. However, AtLAST will be able to efficiently measure percent-level deviations from the dominant thermal SZ effect (see, e.g., Sect.~\ref{sec:test:rtsz} below for a discussion in the context of relativistic SZ corrections) and to swiftly survey wide sky areas at $\sim 1.5-35$~arcsec resolution, thus opening a novel observational window on ICM velocity substructures.

\subsection{Overcoming cluster selection biases}\label{sec:theory:elusive}
It is becoming generally appreciated that X-ray selected clusters offer a biased view of the cluster population \citep{Pacaud2007,Stanek2006,Eckert2011,Planck2011IX,Planck2012I,Maughan2012,Andreon2017,Andreon2019}. This is because, in a given sample, bright clusters are over-represented (see, e.g., \citealt{Mantz2010} for discussion of Malmquist and Eddington biases), whereas those systems fainter-than-average for their mass are underrepresented, if not missing altogether. This bias is difficult to correct because the correction depends on assumptions about the unseen population \citep{Vikhlinin2009,Andreon2017}. On the other hand, SZ-selected cluster samples are generally thought to offer a less biased view and indeed show a larger variety (e.g., in gas content) than X-ray selected samples  (e.g., \citealt{Planck2011IX,Planck2012I}). Comparisons of the X-ray properties of SZ-selected systems (see, e.g., \citealt{chexmate2021}) have highlighted the fact that ICM-based selection biases can depend on the specific morphology \citep{Campitiello2022} or the presence of a dynamically relaxed cool core (the so-called ``cool-core bias''; \citealt{Rossetti2017}). 

However, the selection of clusters via their galaxies (i.e., based on the identification of cluster members) or via gravitational lensing (i.e., based on the effect of the cluster potential on the images of background sources) can provide an observational perspective that is potentially unbiased with respect to the thermodynamic state of the ICM. Although methods based on galaxies can still suffer from significant biases due to contamination and projection effects (e.g., \citealt{Donahue2002,Willis2021}), the fact that they are not dependent on the ICM-specific biases have granted the possibility of unveiling the existence of a variety of clusters at a given mass larger than X-ray or current SZ-based approaches. In particular, the low-surface brightness end of the unveiled new population of clusters is changing our view of galaxy clusters. These are found to introduce significant scatter in many ICM-based mass-observable scaling relations \citep{Andreon2022}, at the very heart of our understanding of cluster physics and broadly used in the context of cluster cosmology. Characterizing such a population of low surface brightness clusters will necessarily require a major leap in the SZ sensitivity with respect to state-of-the-art facilities.

The possibility of performing deep, high angular resolution mapping over wide sky areas offered by AtLAST will allow observers to efficiently detect those clusters that are presently underrepresented in, or entirely missing from, catalogs due to an SZ or X-ray signal inherently fainter than expected from their mass. Indeed, clusters with low X-ray surface brightness tend to have low central values of Compton $y_{\mathrm{t\textsc{sz}}}$, of the order of few $10^{-6}$ (based on \citealt{Andreon2022}), at the very limit of long pointed observations with current single-dish telescopes, when not beyond their effective detection capabilities. In combination with X-ray, strong and weak-lensing data, this will allow for a thorough characterization of their physical and thermodynamic state, and for discriminating between any variation in the inherent properties of the intracluster gas and observational biases induced by any astrophysical processes more or less associated with the specific evolution and physics of the target clusters --- e.g., energetic AGN feedback, recent merger events, low gas fraction, enhanced clustering of millimeter-bright galaxies.

\begin{figure*}[!th]
    \centering
    \includegraphics[width=0.95\textwidth]{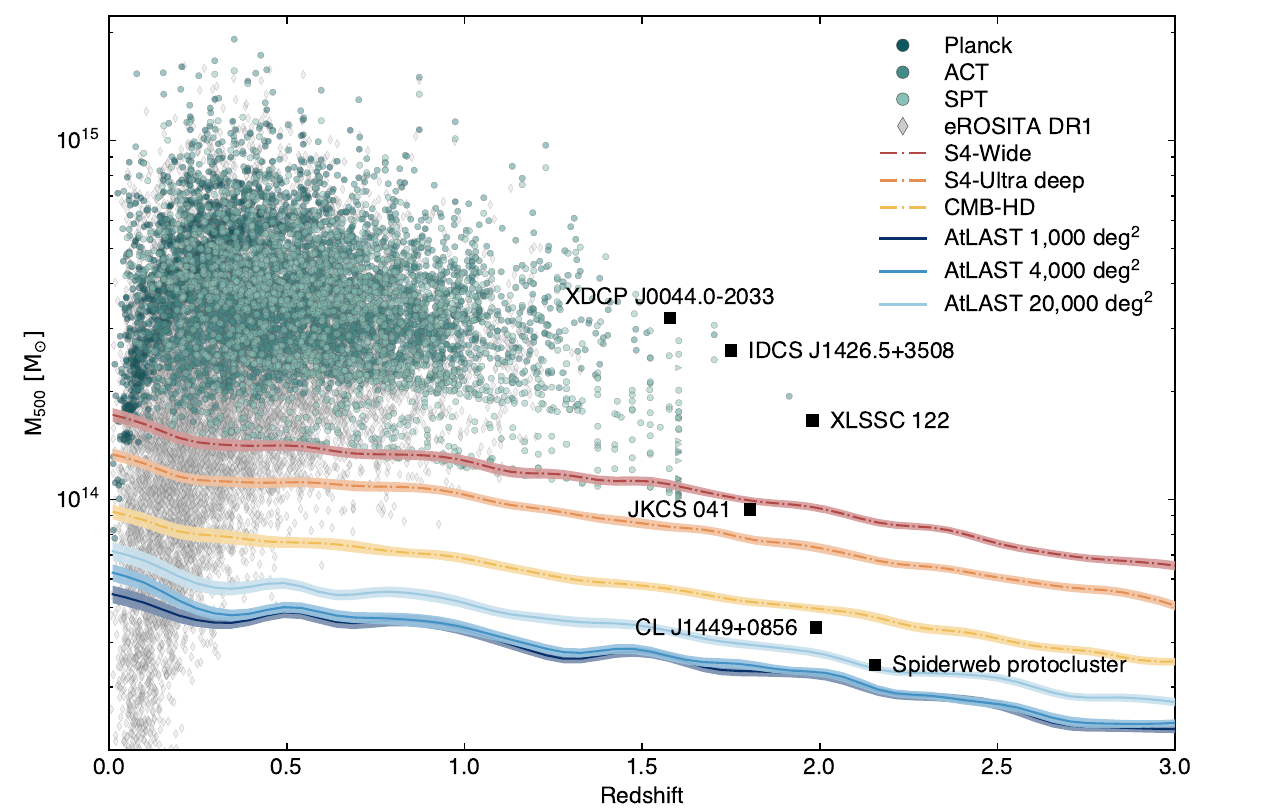}
    \caption{Mass vs.\ redshift detection forecast for AtLAST assuming different survey strategies (covering $1000~\mathrm{deg^2}$, $4000~\mathrm{deg^2}$, and $20000~\mathrm{deg^2}$, respectively, for a fixed survey time of 5 years) in comparison with next-generation wide-field millimeter surveys \citep{CMBS42016,CMBHD2019} and the eROSITA all-sky (X-ray) survey \citep{Bulbul2024}.
    Poisson realizations of the thermal SZ confusion were simulated in the AtLAST survey case, while the other resolution surveys used Gaussian realizations appropriate in the case where the lower resolution and sensitivity limit the ability to surpass the thermal SZ confusion limit. We refer to \citet{Raghunathan2022b} for a more general discussion of the different treatments of the SZ confusion noise. For comparison, we report as green points the clusters from the available SZ survey samples \citep{Planck2016,Bleem2020,Hilton2021,Bleem2023,Kornoelje2025}, as well as relevant high-$z$ clusters from the literature: XDCP~J0044-2033 \citep{Tozzi2015}, IDCS~J1426.5+3508 \citep{Brodwin2016}, JKCS~041 \citep{Andreon2023}, XLSSC~122 \citep{Mantz2018,vanMarrewijk2023}, CL~J1449+0856 \citep{Gobat2019}, and the Spiderweb protocluster \citep{DiMascolo2023}. This figure is adapted from \citet{Raghunathan2022a}. AtLAST will provide an order of magnitude improvement with respect to state-of-the-art SZ facilities, as well as X-ray telescopes in the high-redshift domain. Thanks to the improved mapping speed, AtLAST will also outperform next-generation CMB experiments, allowing a systematic exploration of the low-mass and high-redshift population of cosmic haloes.}
    \label{fig:masszeta}
\end{figure*}

\subsection{Identification and thermodynamic characterization of high-z clusters and protocluster}\label{sec:theory:proto}

Next generation SZ facilities like Simons Observatory \citep[SO;][]{SO2019} and CMB-S4 \citep{CMBS42016} will extend our observational window into the high-$z$ and low-mass realm (see, e.g., \citealt{Raghunathan2022a} and Figure~\ref{fig:masszeta}) of galaxy clusters and protoclusters. Tracing the earliest phases of their evolution will be crucial for constraining the physical origin of the thermal properties of the large-scale structures observed in the nearby Universe. 

Nevertheless, current forecasts estimate that next-generation wide-field surveys \citep{Gardner2023} will detect less than $20\%$ of the most massive (proto)clusters ($M_{200}\lesssim10^{14}~\mathrm{M_{\odot}}$, $z>2$). This is mostly a consequence of the competing impact of inherently low SZ amplitudes (due to low mass, disturbed state, and severe deviations from full gas thermalization and virialization; \citealt{Sereno2021}, \citealt{Bennett2022}, \citealt{Li2023}), the low angular resolution of the facilities, and of the increasing contamination level due to, e.g., enhanced star formation and AGN activity, or possibly due to massive CGM gas and dust reservoirs at high redshift \citep{Lee2024}. And as already broadly discussed in Sect.~\ref{sec:theory:profile}, extreme limitations are also faced in the case of high angular resolution measurements. Clearly, having access to deep, high angular resolution and multi-band SZ observations will allow observers to simultaneously tackle all such issues, making AtLAST the optimal telescope that will definitively shape our perspective on high-$z$ (proto)clusters. To provide a direct comparison of AtLAST's expected capabilities in probing the low-mass and high-$z$ realm with state-of-the-art and planned facilities, we show in Figure~\ref{fig:masszeta} a forecast of the mass-redshift detection threshold for different survey strategies proposed for AtLAST. AtLAST will be capable of improving upon current SZ cluster surveys \citep{Planck2016,Bleem2020,Hilton2021,Bleem2023} almost by an order of magnitude in the lowest detectable mass, practically independently of the redshift. Also in the case of the reference next-generation CMB surveys ––– CMB-S4 \citep{Abazajian2019,Raghunathan2022a}, CMB-HD \citep{CMBHD2019,CMBHD2022} ---, the large collecting area and dense focal-plane arrays will allow AtLAST to achieve a $2-4\times$ improvement in the mass detection limit. This has two key advantages: at low redshift, the possibility of probing the low-mass end of the galaxy cluster and group population would make the SZ effect entirely competitive with respect to X-ray observations (see, for instance, the comparison in Figure~\ref{fig:masszeta} between the eROSITA sample and the AtLAST mass-redshift limits at $z\lesssim0.5$); at high redshift ($z\gtrsim1.5$), this would turn into the possibility of probing all those low-mass haloes that are currently not accessible due to a combination of limited sensitivity and the steep decline of the cosmic halo mass function at high redshift \citep{Cooray2002b,Asgari2023}.

\begin{figure}[!ht]
    \centering
    \includegraphics[clip,trim=0 0 0 0,width=0.475\textwidth]{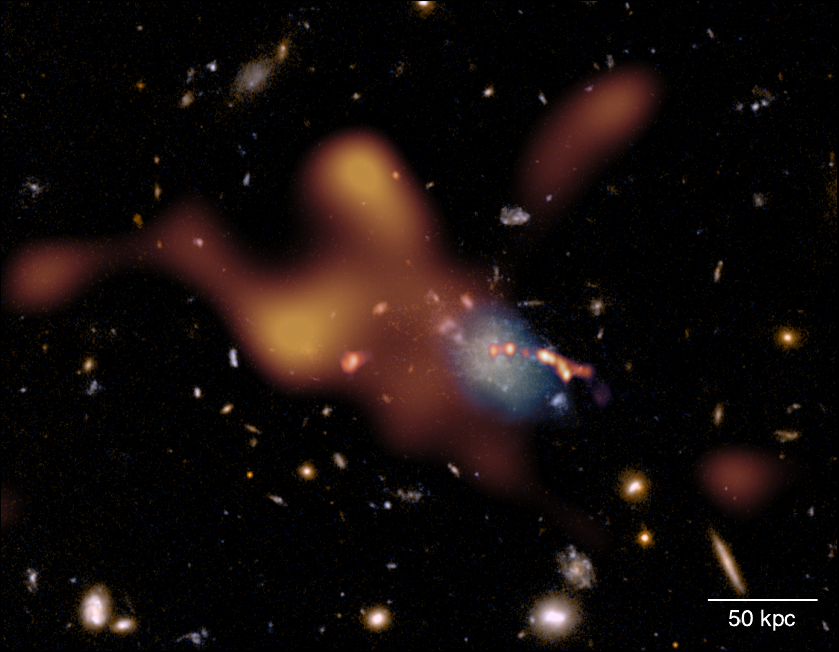}
    \caption{Composite Hubble Space Telescope (HST) image based on ACS/WFC F475W and F814W data of the Spiderweb protocluster field. Overlaid (orange) is the thermal SZ signal from the ICM assembling within the protocluster complex as observed by ALMA over a total of more than $12~\mathrm{h}$ of on-source integration time. In a similar amount of time and with the same spectral tuning, AtLAST will achieve a depth comparable to ALMA, however providing a dramatic improvement of $\sim10^{3}$ in field of view and, thus, in overall mapping speed. At the same time, AtLAST will provide a novel perspective on the reservoirs of cold gas (light blue overlay; \citealt{Emonts2016}) coexisting with the warm/hot phase within protocluster cores (see the CGM science case study by \citet{Lee2024} for a discussion). For comparison, we also include the bright jet of radio emission output from the central galaxy as observed by VLA (the linear east-west feature, shown in red; \citealt{Carilli2022}).
    The present figure is adapted from \citet{DiMascolo2023} and the corresponding \href{https://www.eso.org/public/news/eso2304/}{ESO Press Release eso2304}.}
    \label{fig:spiderweb}
\end{figure}

The correlation of the constraints on the physical and thermodynamic properties of (proto)cluster complexes with the properties of the galactic populations observed within them will further allow for directly linking the evolution of the forming intracluster gas to the multi-phase protocluster environment and its only partially understood impact on galaxy formation and evolution (Figure~\ref{fig:spiderweb}). Current multi-wavelength observations have highlighted that the environmental effects might act on Mpc scales and well beyond the more or less virialized regions within these protocluster galaxy overdensities \citep{Alberts2022}. These studies however rely on the characterization of environmental processing solely from the perspective of protocluster galaxies \citep{Overzier2016}. On the other hand, the wide field, the extreme sensitivity and the capability of AtLAST to trace low density regions thanks to the SZ effect will allow for an efficient imaging of the complex galaxy-environment puzzle with a comprehensive glance of the multi-scale and multi-phase nature of high-$z$ (proto)cluster systems.

\subsection{Impact of AGN feedback and halo heating}\label{sec:theory:AGNfeedback}
Acting in practice as a calorimeter of astrophysical electron populations, the thermal SZ effect can shed light on the interplay of feedback processes and heating of large-scale haloes from galactic to cluster scales. This is particularly relevant in the context of AGN studies, in relation to the specific impact of feedback and AGN-driven outflows in contributing to the heating of cosmic haloes \citep{Fabian2012}. In fact, despite the importance of supermassive black holes (SMBH) in driving the evolution of cosmic structures, we still have a limited understanding of the complex connection between multi-scale physical properties of SMBH and their host galaxies \citep{Gaspari2020}.

Current multi-wavelength observations support a rough duality in the feedback framework \citep{Padovani2017}, with the level of radiative efficiency depending on the specific scenario regulating SMBH accretion \citep{Husemann2018,HlavacekLarrondo2022}. From the perspective of the observational properties of the hot ICM/CGM phase, different feedback models would naturally result in different levels of energy injection and, thus, in deviations from the halo thermal budget expected from virial considerations. At the same time, the strong interaction of winds and jets with the surrounding medium introduces a significant amount of non-thermal support to the overall pressure content --- in the form, e.g., of turbulent motion, buoyantly rising bubbles of extremely hot plasma ($\gtrsim100~\mathrm{keV}$) and associated shock-heated gas cocoons \citep{Pfrommer2005,Abdulla2019,Ehlert2019,Marchegiani2022,OrlowskiScherer2022}. 
All this implies, however, that gaining a detailed view of the thermodynamic properties of the circumgalactic haloes would allow us to obtain better insights into the AGN energetics and improve our feedback models. 

Recently, multiple studies (e.g., \citealt{Moser2022}, \citealt{Chakraborty2023}, \citealt{Grayson2023}) showed that obtaining high angular resolution observations of the thermal SZ effect (in combination with X-ray observations) would allow for constraining the distinct contribution from different feedback models. In fact, measurements of the integrated thermal SZ signal have already been broadly demonstrated to provide an efficient means for probing the evolution of the imprint of feedback on the thermal energy of cosmic structures \citep{Crichton2016,Hall2019,Yang2022}. Concurrently, the first observational studies based on the cross-correlation of the thermal and kinetic SZ signals (e.g., \citealt{Amodeo2021, Schaan2021,Vavagiakis2021,Das2023})  have already shown independent and competitive constraints. Recently, \citet{Coulton2024} demonstrated that the so-called ``patchy screening'' can provide an alternative and highly complementary perspective on feedback mechanisms. 
These are however limited mostly to stacking measurements of arcminute-resolution SZ data, and are thus hampered by systematics deriving from the low angular resolution of the wide-field survey data employed (e.g., the impossibility of cleanly separating first and higher-order halo terms; \citealt{Hill2018,Moser2021,Moser2023,Popik2025}). On the other hand, targeted observations at higher angular resolution currently comprise an extremely small set of high-$z$ quasars \citep{Lacy2019,Brownson2019,Jones2023}. The overall limited sensitivity as well as interferometric effects such as poor $uv$-coverage and the filtering of large scales, however, resulted only in what appear to be low significance detections of the SZ signal in the direction of these systems. 
While these works have been pioneering for high resolution studies, they so far provide little constraining power on the AGN energetics and feedback scenarios. In fact, as demonstrated using numerical predictions for different feedback models \citep{Yang2022}, extending our observational constraints to include a broad range of masses and redshifts, as well as distinguishing between different feedback models, will be highly impractical with current high angular (subarcminute) resolution facilities. Further, beyond the large-scale imprints of feedback on cosmic haloes, it is worth noting that strongly asymmetric outflows from quasars, as well as gas inflows, would result in small-scale distortions of the overall SZ signal due to the localized thermal, kinetic and relativistic SZ contributions (see, e.g., \citealt{Bennett2023}). Similarly, the inflation of cavities by large-scale jets and the consequent generation of shock fronts and turbulent motion would imprint observable deviations in the global SZ signal in the direction of AGN hosts \citep{Ehlert2019}. Having access to sensitive, multi-frequency observations as provided by AtLAST would thus be crucial, on the one hand, for reducing any biases associated with the missing decomposition of the different SZ components to the measured signal as well as any contamination (due to, e.g., millimeter/submillimeter bright emission from the AGN within the studied haloes). On the other hand, it will allow for cleanly dissecting the spectral and morphological features characteristic of the different feedback scenarios.

The SZ signal from AGN-inflated bubbles has instead been robustly detected in one extreme case (MS~0735.6+7421; \citealt{Abdulla2019}, \citealt{OrlowskiScherer2022}). Still, the observations required 10s hours with the current-generation MUSTANG-2 instrument \citep{Dicker2014}, and 100s of hours with the previous-generation CARMA interferometer \citep{Woody2004}, and were each limited to single frequency observations. Since the SZ signal scales as the amount of energy displaced, future observations with current instruments to observe additional, less energetic AGN outbursts could require much more time on the source. As such, this singular example serves largely as a proof-of-principle for further, future resolved studies. We note that some progress will be made in this decade with, e.g., TolTEC \citep{Bryan2018}, though the Large Millimeter Telescope Alfonso Serran (LMT; \citealt{Hughes2010}) was designed to achieve a surface accuracy of $\sim 50~\mu$m (2.5$\times$ worse than AtLAST), and regardless will be limited by the atmospheric transmission to $\nu \lesssim 350~\mathrm{GHz}$ in all but the most exceptional weather \citep[see, e.g., the site comparison in][]{Klaassen2020}. Other single dish facilities delivering similarly high resolution will be limited to similar, if not lower frequencies --- e.g., the 45-meter Nobeyama Radio Observatory (\href{https://www.nro.nao.ac.jp/~nro45mrt/html/index-e.html}{NRO}), the 100-m GBT, the 64-m Sardinia Radio Telescope (SRT; \citealt{Prandoni2017}), and the Institute for Millimetric Radio Astronomy (IRAM) 30-meter telescope ---, while ALMA has difficulty recovering scales larger than $1\arcmin$ in all but its lowest bands (see Section~\ref{sec:technical}).

The reconstruction of the thermal properties of the CGM will have an impact beyond the context of the evolution of the physical processes driving the heating of cosmic haloes. In fact, it will be possible to swiftly build a multi-phase picture of the CGM by concurrently tracing its cold phase along with direct constraints on the otherwise elusive warm/hot constituent --- comprising $\approx80\%$ of the total baryonic material in the CGM overall (e.g., \citealt{Schimek2024}). This is an unparalleled feature of (sub)millimeter measurements, which necessarily require a combination of high spectral and angular resolution, along with the capability of mapping large-scale diffuse signals. AtLAST will be the optimal facility for such a task. For a broader discussion of the importance of multi-phase CGM studies in the context of galaxy formation and evolution, we refer to the companion AtLAST CGM science case study by \citet{Lee2024}.

\subsection{Galaxy cluster outskirts and intercluster structures}\label{sec:theory:whim}

\begin{figure*}
    \vspace{-0pt}
    \begin{center}
    \includegraphics[clip, trim=0mm 0mm 0mm 0mm, width=0.49\textwidth]{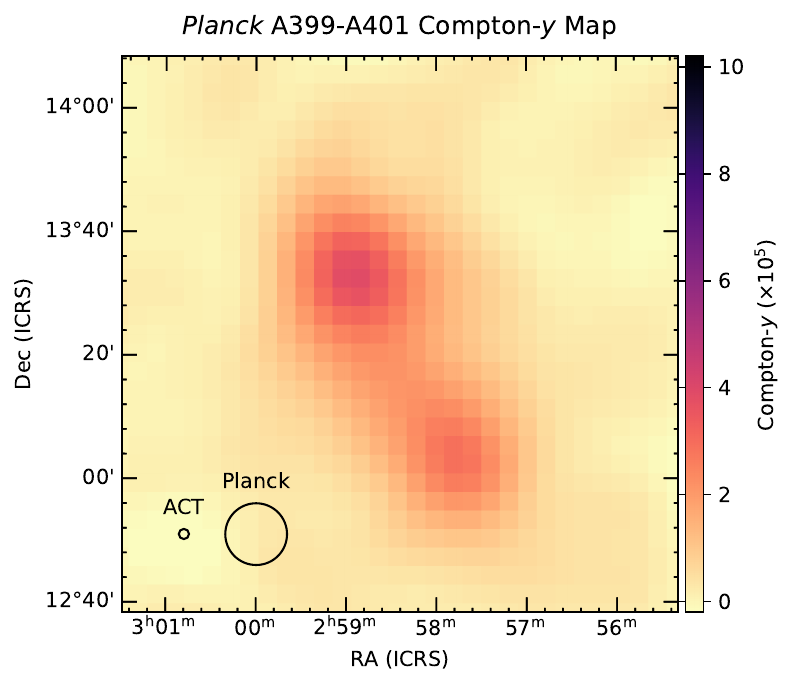}
    \includegraphics[clip, trim=0mm 0mm 0mm 0mm, width=0.49\textwidth]{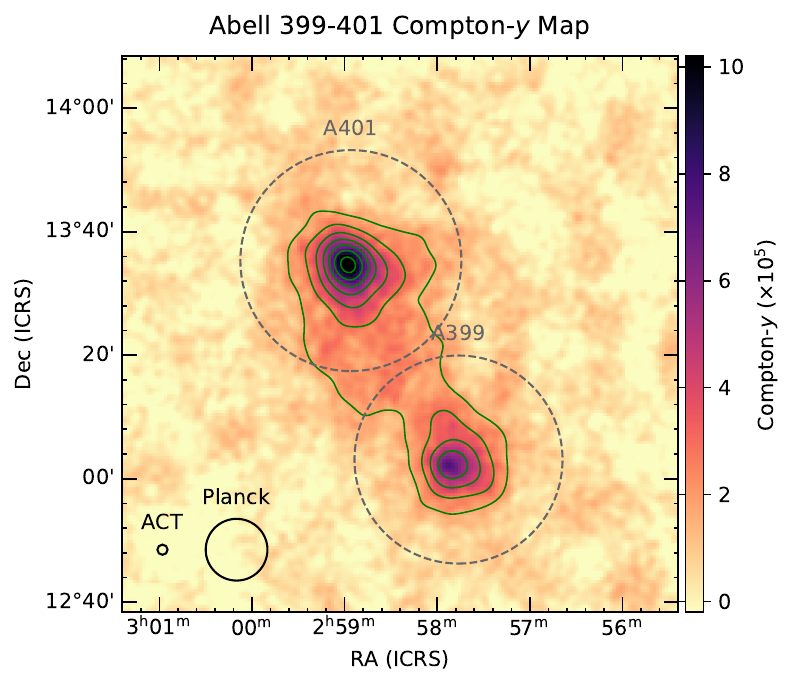}
    \end{center}
    \vspace{-20pt}
    \caption{Comparison of the thermal Compton $y_{\mathrm{t\textsc{sz}}}$ maps produced with \textit{Planck} alone (left, 10\arcmin\ resolution) and by combining \textit{Planck} and ACT (right, 1.7\arcmin\ resolution; we refer to \citealt{Madhavacheril2020} for details on the data combination procedure). When considering the same observing frequency, AtLAST will deliver an $\sim8.3\times$ improvement in resolution (e.g., by achieving beam full-width-half-maximum of $9.83\arcsec$ in the proposed $151~\mathrm{GHz}$ Band 4, compared to the $1.40\arcmin\times1.34\arcmin$ ACT beam at $148~\mathrm{GHz}$), and 69.4$\times$ the instantaneous sensitivity per beam, with respect to that of ACT (and other 6-meter CMB experiments), allowing one to image substructures in intercluster bridges and directly identify and remove source contamination.     
    The figure has been reproduced and adapted with permission from \citet{Hincks2022}.}
    \label{fig:A399-A401}
    \vspace{-0pt}
\end{figure*} 

A significant portion of the baryonic content of the Universe at $z\lesssim3$ is expected to lie well beyond the virial boundaries of cosmic structures \citep{Cen1999}. This diffuse ``warm-hot intergalactic medium'' (WHIM) is expected to have temperatures $T_{\mathrm{e}} \approx 10^5 - 10^7~\mathrm{K}$, largely invisible at optical wavelengths and generally too low in temperature for all but the deepest X-ray observations, often being limited to line-of-sight absorption studies \citep{Nicastro2018}. Obtaining a detailed view of the large-scale WHIM is however crucial. Accurately constraining the actual amount of matter constituting the WHIM will provide fundamental information on the ``missing baryons'' budget associated to this specific phase of the filamentary IGM \citep[e.g.,][]{Shull2012}. This will be connected to the specific mechanisms driving the heating of large-scale structure on cosmological scales: on the one hand, matter inflows and mergers along large-scale filaments driving strong accretion and virialization shocks (\citealt{Baxter2021,Anbajagane2022,Anbajagane2023}; see also Sect.~\ref{sec:theory:profile}); on the other hand, the impact of feedback processes and of the environmental pre-processing of galaxies \citep[e.g.,][]{Fujita2004,Alberts2022}.

To date, the identification and characterization of the physical properties of the filamentary WHIM has been performed mostly through stacked SZ and/or X-ray measurements (e.g., \citealt{deGraaff2019,Tanimura2019,Singari2020,Tanimura2020,Tanimura2022}), and is often dominated by the hottest extremes of the range of temperatures expected for the WHIM (see \citealt{Lokken2023} for discussion). Recently, direct SZ imaging of a nearby intercluster bridge was presented in \cite{Hincks2022}, which used the combination of ACT+\textit{Planck} data to reveal details at a much higher spatial dynamic range than the previous results using \textit{Planck} alone. The results are shown in Figure~\ref{fig:A399-A401}. Among the key aspects of this work is the fact that it has served as an invaluable pathfinder for the imaging of IGM substructures via the SZ effect. This is a direct consequence of the key advantage provided by observations of the thermal and kinetic SZ effects --- which scale linearly with the electron density (see Sect.~\ref{sec:sz}) --- over X-ray surface brightness measurements --- which scale as the density squared --- in the broader context of the study of low-density environments like large-scale filaments and cluster outskirts.

Still, observations like the one by \citet{Hincks2022} clearly highlight how the possibility of proving IGM filaments at modestly high ($\sim 6\times$) resolution is still limited to nearby ($z\approx0.05$) massive clusters. Deep maps with AtLAST will instead allow for improved spatial dynamic range and higher fidelity, enabling such studies for many more clusters going to both higher redshifts and lower mass regimes. 
Thanks to its sensitivity and to the possibility of recovering large scales over extremely wide fields, AtLAST will provide the ideal tool for searching for the presence of the SZ effect in accreting and unbound intergalactic gas surrounding the virialized volume of clusters and groups (we refer to Figure~\ref{fig:tsz-ksz} for a simulated perspective on AtLAST's capability of probing cluster outskirts). In particular, this will allow for routinely exploring intercluster structures in a large number of cluster pairs without the need for time demanding observations. For instance, it will be possible to achieve the same Compton $y_{\mathrm{t\textsc{sz}}}$ (or surface brightness) sensitivity as in the observation of the A399-A401 observations by \citeauthor{Hincks2022}\ (\citeyear{Hincks2022}; see also Figure~\ref{fig:A399-A401}) in less than $\sim10~\mathrm{h}$ of integration, but with better spectral coverage and an order of magnitude improvement in the angular resolution. Clearly, such a comparison is only based on sensitivity considerations and only holds  in the case of filaments extending over scales comparable to AtLAST's instantaneous field of view (as in the reported case of A399-A401). Large-scale filtering resulting from either sub-optimal scanning strategies or the limited capability of disentangling any large-scale signals from atmospheric common-mode noise fluctuations can severely impact the possibility of detecting intergalactic filaments extending beyond degree scales (see, however, the discussion in Sect.~\ref{sec:technical:instrument} below). On the other hand, we can consider as a rough lower limit of the expected amplitude for large-scale filaments the results from previous stacking experiments on intergalactic gas. For instance, \citet{deGraaff2019} provide estimates of the average SZ signal to have amplitudes in Compton $y_{\mathrm{t\textsc{sz}}}$ unit of $\lesssim10^{-8}$, corresponding to a maximum amplitude of the thermal SZ signal of $\gtrsim-64~\mathrm{nJy~beam^{-1}}$ for the decrement, and $\lesssim 9.5~\mathrm{nJy~beam^{-1}}$ for the increment. Although impractical for performing any direct imaging of WHIM between and around individual galaxies, the extreme observing speed of AtLAST will allow for extending the stacking constraints to higher redshift and smaller physical scales, providing a resolved evolutionary perspective on the hot phase of the cosmic web and the processes driving their thermal properties. 
Similarly, the combination of the broad spectral coverage with the arcsecond-level angular resolution will allow for reducing contamination from inter-filamentary structures, either in the form of a Poisson SZ noise from small-mass haloes within the filaments themselves or of bright (sub)millimeter emission from individual galaxies. At the same time, this will provide the means for directly inferring robust temperature constraints (currently representing the main limitation for using the SZ effect for determining the overall contribution of WHIM to the missing baryon budget). In this regard, we refer the reader to the discussion about temperature reconstruction using measurements of the relativistic corrections to thermal SZ effect (Sect.~\ref{sec:sz:rsz} and Sect.~\ref{sec:theory:rtSZ}).

\subsection{Beyond SZ intensity measurements}
In the discussion above, we have focused solely on the novel observational opportunities that will be opened by AtLAST's capabilities in observing the SZ effects and, in general, (sub)millimeter sky in total intensity. As summarized in the AtLAST Science Overview by \citet{Booth2024a}, other science case studies have clearly highlighted how the integration of polarimetric capabilities will allow for opening up to major observational advances --- for instance, for the study of the magnetic field in flaring regions of our Sun \citep{Wedemeyer2024} or embedded in the interstellar medium of the Milky Way \citep{Klaassen2024} and nearby galaxies \citep{Liu2024}.

In the context of the SZ effect, the high polarization sensitivity that will be offered by AtLAST could provide the means for gaining an unprecedented perspective on the polarized counterparts to the dominant SZ terms introduced in Sect.~\ref{sec:sz} and \ref{sec:theory}. Any anisotropy in the radiation field  --- e.g., the quadrupole term of CMB radiation \citep{Sazonov1999}, analogous the generation of primordial CMB polarization \citep{Hu1997,Kamionkowski1997} --- or in the scattering medium --- e.g., in the case of clusters in transverse motion on the plane of the sky \citep{Sunyaev1980,Sazonov1999} or in the presence of local pressure anisotropies \citep{Khabibullin2018} --- induce characteristic polarization features that carry key information on the collisional nature of ICM, the velocity structure of haloes and large-scale structures, or on the primordial CMB anisotropies (see \citealt{Khabibullin2018} and \citealt{Lee2024SZ} for detailed discussions; we further refer to \citealt{Mroczkowski2019} for a general review). In fact, as summarised in \citet{Khabibullin2018}, the degree of polarization expected for any polarized SZ component falls below the $\sim10^{-8}$ level, corresponding to a surface brightness $\lesssim 10~\mathrm{nK}$. In turn, directly constraining any polarized contributions to the SZ effect will represent a key challenge in terms of overall polarization sensitivity, decontamination from foreground and background sources, as well as overall polarization accuracy and control of any cross-polarization leakages \citep{Puddu2024,Mroczkowski2025}. Properly forecasting the possibility of measuring the faint polarized SZ effect will thus require a dedicated effort that could include, along with the associated astrophysical effects, any bias associated with specific choices in instrument and optical design for AtLAST, and we leave this for future work.

\section{Technical justification}\label{sec:technical}
The challenging nature of the proposed exploration of the thermal properties of the multi-scale realm of cosmic structure discussed above --- from intergalactic filaments, to the ICM in forming systems and the CGM surrounding individual galaxies --- will necessarily require a major leap in our observational capabilities. Here, we provide a brief list of the technical requirements that will be key for pursuing the aforementioned science drivers. 
\begin{itemize}[leftmargin=*,topsep=2pt,itemsep=2pt,parsep=2pt]
    \item \textbf{Broad spectral coverage.} The different components of the SZ effect exhibit different, well-defined spectral signatures (Sect.~\ref{sec:sz}). It is therefore to disentangle the different SZ contributions from each other, as well as from any contaminating signal, given sufficiently broad spectral coverage. Crucial for achieving an optimal spectral separation of the SZ effects is an extensive, multi-band coverage of the entire relevant frequency range accessible by AtLAST. 

    \item \textbf{Few arcsecond-level angular resolution.} This will be essential to constrain the thermodynamic properties of the ICM and CGM down to their cores, along with any small-scale structures associated to the tumultuous evolution of galaxy clusters (Sect.~\ref{sec:theory:profile}) and galaxies (Sect.~\ref{sec:theory:AGNfeedback}). At the same time, it will mitigate the contamination due to the emission at (sub)millimeter wavelengths from compact sources associated with the massive cluster haloes or populating foreground and background fields, allowing for a more effective extraction of the SZ signals.
    
    \item \textbf{Wide-field capabilities,} aiming at achieving an instantaneous degree-square field of view. This will provide the means for probing the diffuse SZ signal from the warm/hot gas within cluster outskirts, (several) Mpc-intergalactic filaments, and protocluster complexes (Sect.~\ref{sec:theory:proto} and \ref{sec:theory:whim}), while allowing for efficiently surveying the (sub)millimeter sky in search of the faint end of the cluster population (Sect.~\ref{sec:theory:elusive}).
    
    \item \textbf{Sensitivity.} Probing inherently faint and diffuse signals -- as in the case of IGM filaments (Sect.~\ref{sec:theory:whim}) or the relativistic and kinetic SZ terms (Sect.~\ref{sec:theory:rtSZ} and \ref{sec:theory:ksz}) -- will require an improvement in the noise performance well beyond the reach of state-of-the-art and any forthcoming (sub)millimeter facilities. As discussed in the previous section, we specifically aim at achieving noise levels in units of Compton $y$ $\lesssim 5\times 10^{-7}$.    
\end{itemize}
In this section, we will expand upon such requirements and identify the critical aspects for pursuing the proposed SZ studies with AtLAST.

\subsection{Overview of the instrumental requirements}\label{sec:technical:instrument}
The field of view of a (sub)millimeter telescope represents a key parameter in the context of SZ science.
Current large single-dish telescopes lose signals on scales larger ($\approx2-6\arcmin$; see, e.g., \citealt{Adam2015,Ruppin2018,Ricci2020,Romero2020,Okabe2021,Andreon2023,MunozEcheverria2023}) than the instantaneous fields of view of their respective high-resolution instruments --- e.g., MUSTANG-2 \citep{Dicker2014}, NIKA2 \citep{Adam2018}, TolTEC \citep{Bryan2018}, MISTRAL \citep{Battistelli2024,Paiella2024} --- where much of the most interesting, faint target SZ signals exist. We note that continuum observations using the 12-meter antennas in the ALMA Total Power Array (TPA; \citealt{Iguchi2009}) suffer even more egregiously from being unable to remove atmospheric contamination due to their limited fields of view. They also suffer from the poor mapping speeds associated with single beam observations, and from relatively small collecting areas. The issue associated with large-scale filtering is arguably more restrictive in the case of interferometric observations, which generally feature maximum recoverable scales that fall within the sub-arcminute regime (e.g., ALMA Bands 4-10; we refer to the \href{https://almascience.eso.org/proposing/technical-handbook}{ALMA Technical Handbook} for further details).  
So far, instruments with much larger instantaneous fields of view, which are therefore better able to recover larger scales, have been employed in the context of CMB/SZ survey experiments like the Atacama Cosmology Telescope (ACT; \citealt{ACT2011}), the South Pole Telescope (SPT; \citealt{SPT2011}), SO (\citealt{SO2019}), CMB-S4 \citep{CMBS42016}, CCAT (Fred Young Submillimeter Telescope or FYST; \citealt{CCAT2023}). Still, such survey telescopes universally feature small apertures ($\leq 10$-m)  with correspondingly limited collecting areas, resulting in arcminute-level angular resolution and poor source sensitivity, which leaves these telescopes largely unsuitable for exploring the small-scale morphologies of galaxy clusters and protoclusters (apart from a few systems in the nearby Universe). With its unparalleled combination of high angular resolution and field-of-view, AtLAST will thus be ideally positioned to fill the gap between current and future wide-field survey facilities and high-resolution single-dish and interferometric telescopes, finally opening a (sub)millimeter perspective on the cosmic structures and their multi-scale properties.

In general, larger scales are difficult to recover due to the large, and largely common mode, atmospheric signal which dominates. A field of view of reduced size requires a tailored observational strategy and data reduction pipeline to mitigate signal loss at large scales.
Nevertheless, even in such a case, the recovery of astrophysical information beyond the maximum recoverable scales of such facilities would still be severely hampered. This would critically affect many of the proposed science goals, particularly those requiring both wide field of views and extended recoverable scales. Intergalactic filaments are in fact expected to extend over tens of Mpc (e.g., \citealt{GallarragaEspinosa2020}) and, thus, extending over degree-scales in the case of nearby superclusters \citep{Ghirardini2021b}. Similar physical extents are observed also in the case of high-$z$ protocluster complexes ($\lesssim20$ arcmin; see, e.g., \citealt{Matsuda2005,Cantalupo2019,Hill2020,Jin2021}). And as shown in Figures.~\ref{fig:szprof} and \ref{fig:tsz-ksz}, effectively probing the distribution of the ICM thermodynamic properties out to the cluster outskirts requires mapping the SZ signal beyond $\sim1~\mathrm{deg}$ in cluster-centric distance. Therefore, the capability of gaining instantaneous observations of structures extending from few arcminutes up to degree scales will represent a crucial benefit of AtLAST compared to state-of-the-art and future telescopes covering the same observational windows. We note that the expected large-scale performance of AtLAST is currently based on the extrapolation of the results from past and current small-aperture survey telescopes and high-resolution facilities. In fact, as mentioned above, the main single-dish (sub)millimeter facilities currently used for pursuing resolved SZ science have been directly proven to be capable of probing signal on spatial scales comparable to their field of view. This is the case, for instance, of the NIKA \citep{Adam2015} and NIKA2 \citep{Ruppin2018,Keruzore2020,MunozEcheverria2023} instruments, whose effective spatial transfer functions are characterized by large-scale half-power widths even $\sim2\times$ larger than their fields of view ($1.9\arcmin$ and $6.5\arcmin$, respectively). As mentioned above, similar results are obtained also in the case of small-aperture survey telescopes --- with, e.g., ACT reporting 1\% signal losses for $\ell\simeq400$ \citep{ACT2025}, roughly matching its $22\arcmin\times26\arcmin$ field of view \citep{ACT2011}.  And at the same time, \citet{Romero2020} demonstrated how proper modelling of atmospheric noise contributions on field-of-view scales could lead to significant enhancements in the effective sensitivity to, and our ability to recover, large-scale spatial modes. 

It is crucial to note that present-day pathfinders for AtLAST, such as the 100-m GBT or IRAM 30-m, operate in less optimal sites, under atmospheric conditions that are significantly worse than those available on the Llano de Chajnantor in the Atacama \citep[e.g.][]{Klaassen2020}.
Therefore, while they have exceeded expectations in recovering scales comparable to or larger than their instantaneous fields of view, we expect AtLAST to provide even better performance in terms of atmospheric noise mitigation, large-scale signal recovery, and overall mapping efficiency. The detailed characterization of the expected spatial transfer function of AtLAST and the impact on this of different atmospheric conditions, mapping strategies, spectral bands, and detector array configurations will be explored in a future work. Preliminary tests on a reduced configuration are however already reported by \citet{vanMarrewijk2024b}, showing results that are consistent with those reported for existing (sub)millimeter facilities.


\begin{figure*}[!ht]
    \centering
    \includegraphics[clip,trim=0 8pt 0 16pt,width=0.98\textwidth]{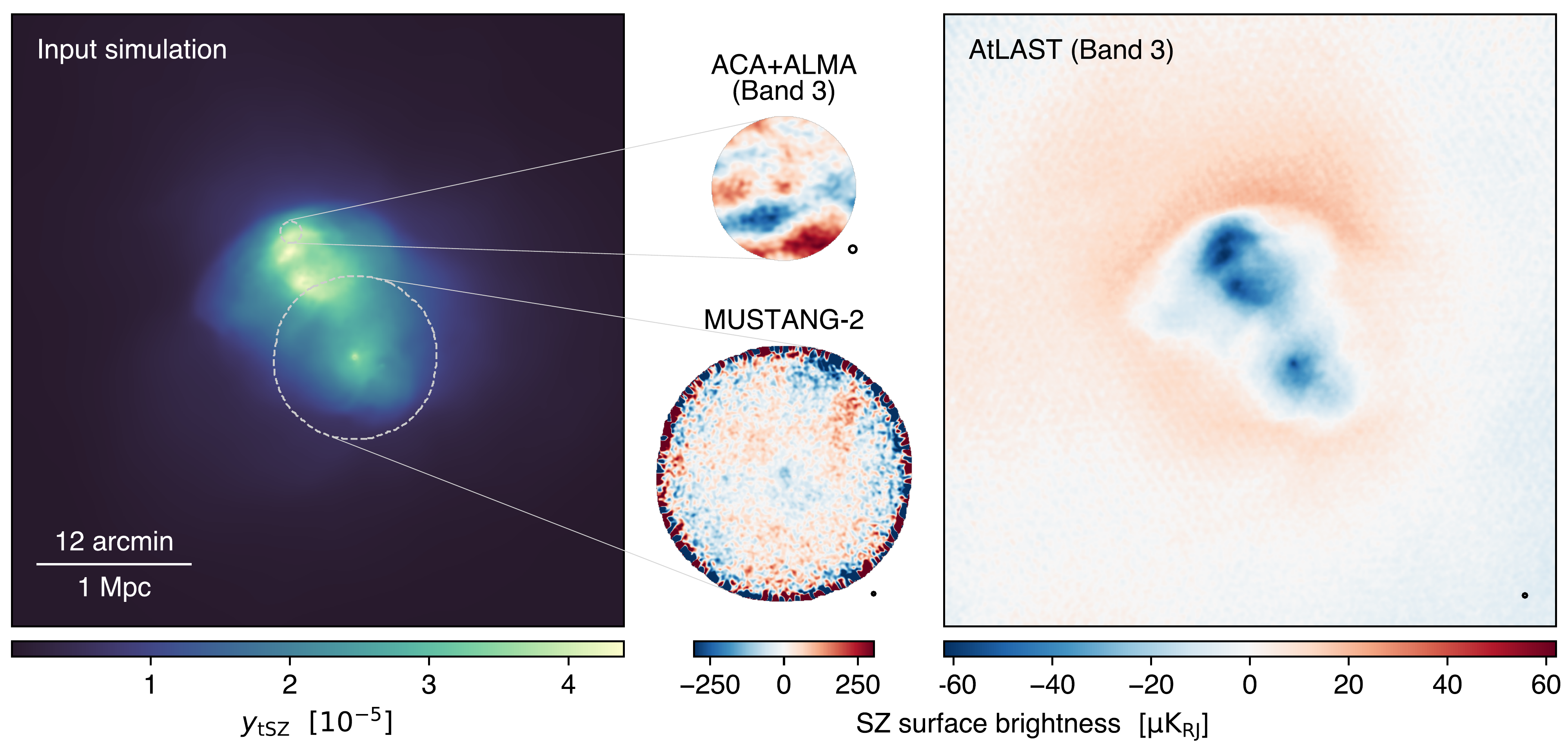}
    \caption{Compton $y_{\mathrm{t\textsc{sz}}}$ (thermal SZ signal) distribution for a simulated nearby galaxy cluster ($M_{\mathrm{500}}=1.28\times10^{15}~\mathrm{M_{\odot}}$, $z=0.0688$; left) as observed by ALMA+ACA in Band~3 (top center), MUSTANG-2 (bottom center), and by AtLAST in ALMA Band~3 (right). The ALMA+ACA and MUSTANG-2 observations were both generated at $90~\mathrm{GHz}$, the nominal bandpass center of the MUSTANG-2 receiver \citep{Dicker2014}. The respective beams are shown in the bottom right corner of each panel. The input simulation is extracted from the Dianoga cosmological simulation suite \citep{Rasia2015,Bassini2020}. Overlaid as dashed white circles are the ACA+ALMA and MUSTANG-2 footprints. We note that the respective panels on the central column are scaled up arbitrarily with the goal of highlighting any observed features, and do not reflect the relative angular sizes of the fields. For all cases, we consider an on-source time of 8 hours. The mock AtLAST and MUSTANG-2 observations are generated using the \texttt{maria} simulation tool (see \citealt{vanMarrewijk2024b} for details), assuming an AtLAST setup with the minimal detector counts of $50\,000$ (Sect.~\ref{sec:test:det}). For ACA+ALMA, we employ the \texttt{simobserve} task part of the Common Astronomy Software Applications (CASA; \citealt{CASA2022}). All the simulated outputs are in $\mathrm{\mu K_{\textsc{rj}}}$ brightness temperature units, with the ALMA and MUSTANG-2 maps shown using the same color scale. Despite the fact the different images are generated adopting a minimal simulation setup (e.g., do not include any foreground and background contamination terms beyond the atmospheric noise), this figure clearly demonstrates the key advantage provided by AtLAST's extended spatial dynamic range and sensitivity for probing the SZ effect from galaxy clusters.}
    \label{fig:mocks}
\end{figure*}

Multi-band observations are also critical to suppress/mitigate non-SZ signals below the detection threshold, making wide spectral coverage essential for many of the science goals detailed above. Current high resolution facilities on large telescopes have at most three bands, and are limited to relatively low-frequency observations  --- e.g.\ $\leq350$~GHz for the LMT \citep{Hughes2010}, $\leq 270$~GHz for the IRAM 30-meter telescope, and $\leq 115$~GHz for the 100-meter GBT \citep{White2022}, the 64-meter SRT \citep{Prandoni2017}, or any potential single dish component of the ngVLA \citep{Selina2018}. This implies that any current or next-generation facilities will provide limited constraining power for the relativistic and kinetic SZ, as well as contamination from the cosmic infrared background or diffuse dust contamination.

Most foregrounds should be spatially distinguishable from the SZ signal. However there may be a spatially coincident large-scale dust component originating from within clusters themselves (e.g., \citealt{Erler2018}) which makes at least two bands in the range $400-900~\mathrm{GHz}$ indispensable to trace the Rayleigh-Jeans tail of the dust spectral energy distribution and to mitigate biases in the SZ spectral modeling. 
An additional band at $\approx$\,1200\,GHz would be even more helpful to resolve degeneracies between dust and SZ signals, but this is precluded by the severe reduction in atmospheric transmission. 

To meet the observational requirements for pursuing the proposed scientific goals (Sect.~\ref{sec:theory}), we perceive the most salient instrumentation requirements to be the ability to achieve high continuum mapping speeds over large areas and in multiple bands. This would convert into the key demand of densely filling the telescope focal plane with a large count of multi-frequency detectors. In this regard, the different detectors (e.g., transition edge sensor bolometers or kinetic inductance detectors) already in use in current state-of-the-art continuum cameras --- such as MUSTANG-2 \citep{Dicker2014}, NIKA2 \citep{Adam2018}, TolTEC \citep{Bryan2018} or MISTRAL \citep{Paiella2024} ---have already demonstrated a high technical readiness level, providing background-limited performance in the (sub)submillimeter the possibility of being read out in large numbers (tens-to-hundreds of thousands, as noted in \citealt{Klaassen2020}) through frequency multiplexing, allowing the construction of large imaging arrays. We further refer to the \href{https://www.atlast.uio.no/memo-series/memo-public/instrumentationwgmemo4_29feb2024.pdf}{AtLAST Memo 4} for details on the expected instrumental specifications for AtLAST.

To illustrate the observational capabilities of the proposed AtLAST continuum setup, we generate mock observations using the \texttt{maria} simulation library \citep{vanMarrewijk2024b} and consider a simulated galaxy cluster extracted from the Dianoga hydrodynamical cosmological simulations \citep{Rasia2015,Bassini2020} as input. The results are shown in Figure~\ref{fig:mocks}. For comparison, we include simulated observations performed with MUSTANG-2 and jointly with ALMA and the 7-m Atacama Compact Array (ACA; \citealt{Iguchi2009}). The clear result is the superior capability of AtLAST in recovering spatial features over a broad range of scales at high significance, while MUSTANG-2 and ACA+ALMA suffer from limited sensitivity and significant large-scale filtering, respectively. We note that, in this test, we are considering only single-band observations at the same frequency to facilitate the comparison. Although ALMA Band 1 offers an improved sensitivity, spatial dynamic range, and field of view compared to Band 3, it still provides a limited sampling of large-scale SZ structures (with a maximum recoverable scale $\mathrm{MRS}\lesssim1.20\arcmin$ when ALMA is in its most compact configuration).

\subsection{Optimizing the spectral setup}\label{sec:test:bands}
As mentioned broadly in Sect.~\ref{sec:theory} and discussed in the introduction to this section, among the critical aspects for performing a robust reconstruction of the SZ effect is the requirement of cleanly separating the multiple spectral components of the SZ signal from contaminating sources. From a technical point of view, this converts to maximizing the spectral coverage while requesting maximum sensitivity (i.e., lowest noise root-mean-square) for each of the bands. Given the deteriorating atmospheric transmission when moving to higher frequencies, this is not obtained by trivially expanding the effective bandwidth arbitrarily. At the same time, we would like to consider a minimum setup in order not to result in an over-sampling of the target spectral range. If a finer spectral sampling would allow for improved capabilities of AtLAST in terms of component separation, each band would entail additional demands in terms of instrument complexity and overall costs.

\begin{figure}[!ht]
    \centering
    \includegraphics[clip,trim=0 8pt 0 6pt,width=0.49\textwidth]{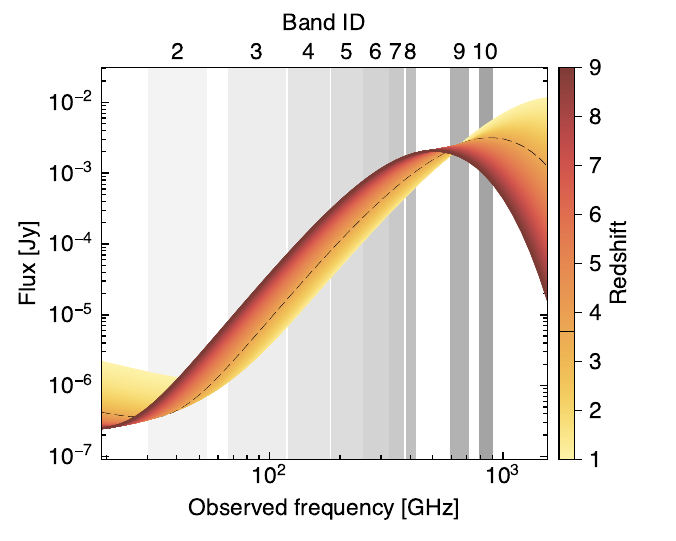}
    \caption{The high-frequency bands (Band 8-10) will be crucial for optimally sampling the peak of the dust continuum emission from individual high-$z$ background galaxies. Along with inferring the physical properties of their dust content, this will be crucial for minimizing the contamination of the SZ signal due to cospatial dusty components (see also Sect.~\ref{sec:test:rtsz}). As a reference, we show here model emission for star-forming galaxies at varying redshift. Here, the dust contribution is based on the $z$-dependent dust temperature model from \citet{Sommovigo2022} for a dust mass $M_{\mathrm{d}}=10^8~\mathrm{M_{\odot}}$ (consistent with the galaxy REBELS sample; \citealt{Bouwens2022}). The low-frequency radio component reproduces the radio model from \citet{Delvecchio2021}, assuming an infrared-to-radio luminosity ratio of $q_{IR}=2.646$. The dashed line denotes the lowest redshift at which the dust emission peak falls within the AtLAST spectral range ($z\simeq3.60$), thus allowing for an accurate modelling of the corresponding spectrum}.
    \label{fig:dusty}
\end{figure}

\begin{figure}[!ht]
    \centering
    \includegraphics[width=0.49\textwidth]{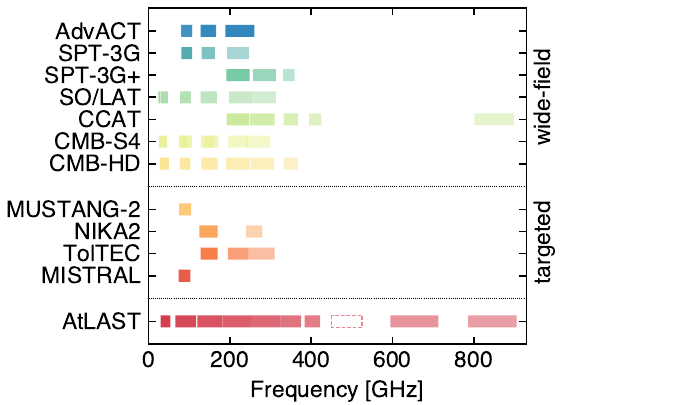}
    \caption{Comparison of the expected AtLAST bands (see Table~\ref{tab:freq_sens_beam} and Fig.~\ref{fig:sz_spectrum}) with the spectral coverage for the main previous, existing, or planned (sub)millimeter facilities suitable for dedicated SZ studies. The listed wide-field survey instruments are: Advanced ACT receivers \citep{Thornton2016}; SPT-3G \citep{Sobrin2018,Sobrin2022} and the corresponding SPT-3G+ extensions \citep{SPT3Gp2022}; SO Large Aperture Telescope (LAT) receivers \citep{Zhu2021,Bhandarkar2025}; CCAT Prime-Cam \citep{Vavagiakis2018,Vavagiakis2022}; CMB-S4 LAT cameras \citep{Gallardo2022,Gallardo2024b}; CMB-HD receivers \citep{CMBHD2019}. The reported high-angular resolution instruments employed for targeted SZ observations are, instead: GBT/MUSTANG-2 \citep{Dicker2014}, IRAM/NIKA2 \citep{Adam2018}, LST/TolTEC \citep{Bryan2018}, and SRT/MISTRAL \citep{Paiella2024}. The AtLAST band shown using dashed contours corresponds to the $\sim500~\mathrm{GHz}$ spectral window, excluded in our forecasts due to the limited sensitivity resulting from the low atmospheric transmission and narrow effective bandwidth. Even when excluding such a band, AtLAST will provide a uniquely wide spectral coverage.}
    \label{fig:band_compare}
\end{figure}

A summary of the selected bands, specifically optimized to minimize the output noise root-mean-square level per given integration time, is provided in Table~\ref{tab:freq_sens_beam}. Fig.~\ref{fig:band_compare} provides a direct comparison of the spectral coverage proposed for AtLAST and of other (sub)millimeter facilities, clearly highlighting how AtLAST will be capable of surpassing other telescopes by accessing a uniquely wide frequency range. Our low-frequency set ($\lesssim 300~\mathrm{GHz}$) extend upon the multi-band set-up proposed for CMB-S4 \citep{CMBS42016}, shown in forecasts to provide an optimal suppression of the contribution from astrophysical foregrounds and backgrounds \citep{Abazajian2019}. Motivated by the expected coverage of the $\lesssim 30~\mathrm{GHz}$ range by future radio facilities (e.g.\ SKA, ngVLA), we decided not to include the synchrotron-specific $20~\mathrm{GHz}$ band. Still, the spectral setup proposed over the millimeter wavelength range will be useful well beyond the SZ science. Having access to sensitive and high resolution continuum measurements will offer, for instance, the opportunity of probing the high-frequency regime of the free-free and synchrotron emission from individual galaxies to get dust-unbiased perspective on star formation (\citealt{Condon1992,Murphy2011,Murphy2012}; we further refer to the AtLAST science case by \citealt{Liu2024}), accessing key information on the coronal origin of the millimeter-wave signature of radio quiet AGN \citep{Behar2018,Panessa2019,delPalacio2025}, or shedding light on the elusive Anomalous Microwave Emission (AME) and its potential association with spinning and magnetized interstellar dust \citep{Dickinson2018}.

On the other hand, given the centrality of the high-frequency ($\gtrsim500~\mathrm{GHz}$) for maximizing AtLAST's capability of separating different SZ components and the signal from contaminating sources (see Sect.~\ref{sec:test:rtsz}), we extend the overall spectral coverage beyond $300~\mathrm{GHz}$ to include four additional bands up to $900~\mathrm{GHz}$. Compared to FYST's choice of survey bands \citep{CCAT2023}, our choice will allow one to better sample the high-frequency end of AtLAST's spectral range and, in turn, to gain a better handle on the relativistic SZ effect and on the dust emission from any extended intracluster reservoirs or within individual galaxies (either associated with the clusters, or in background and foreground fields). In this regard, we show in Figure~\ref{fig:dusty} the redshift-dependent spectral model for dusty star-forming galaxies (based on the results by \citealt{Sommovigo2022}). By probing the full Rayleigh-Jeans tail of the dust spectral energy distribution, it will be possible, on the one hand, to gain thorough insight into the dust content of galaxies (and, thus, star-formation processes; \citealt{Casey2014,Schneider2024}) across a wide range of redshift. On the other, it will provide the means for mitigating the dust contamination of the high-frequency end of the SZ spectrum. We further refer to \citealt{vanKampen2024} for a direct comparison of the large-scale distribution of submillimeter bright sources observed by the arcminute-resolution ACT and AtLAST. 

It is worth highlighting that the high angular resolution provided by AtLAST will be a key aspect in complementing and significantly enhancing the possibility of cleanly subtracting the emission from compact foreground/background sources, allowing for their clean separation from the SZ effect by means of a full multi-frequency and multi-scale approach. We here note, though, that an accurate spectral modelling of the multi-faceted contributions to the SZ effect (and of their contaminants) can not forego the inclusion of beam chromaticity effects in the analysis. The frequency-dependent variation of the angular resolution both for different bands and across the broad passband of a given spectral window can in fact introduce spurious biases in the measured spectral properties of any target continuum emission. A recent forecast analysis by \citet{Giardiello2025} has however highlighted how a simple forward modelling treatment of beam chromaticity can significantly mitigate its impact on CMB power spectrum studies. Although we envision adopting a similar approach also in the case of AtLAST SZ measurements, we postpone to future dedicated works the exploration of pratical solutions for allowing for a spatio-spectral modelling of high dynamic range observations.

Finally, we note that we are currently investigating the possibility of integrating an additional band covering the $\sim500~\mathrm{GHz}$ atmospheric window, but the low transmission and limited fractional bandwidth are expected to limit the effectiveness of such an addition in terms of an increase of the overall SZ sensitivity. However, we emphasize that AtLAST coverage of the ALMA Band~8 frequencies up to $\nu= 492~$GHz would be fundamental for other application in the context of AtLAST science. We refer the interested readers to the companion AtLAST case studies by \citet{Lee2024} and \citet{Liu2024}. 

\begin{sidewaystable*}
	\centering
    \rule{\textwidth}{0.4pt}\vspace{-6pt}
    \caption{Frequencies, sensitivities and beam sizes for a large SZ survey carried out with AtLAST. The sensitivity levels are computed assuming standard values for the weather conditions (2\textsuperscript{nd} octile) and elevation ($\alpha=45\deg$). To account for the variation of the atmospheric transmittance and of the system temperature across a given band, the sensitivity is obtained by considering the in-band inverse root mean square average of the frequency-dependent system equivalent flux densities provided by the AtLAST sensitivity calculator (\citealt{senscalc}; see also the \href{https://atlast-sensitivity-calculator.readthedocs.io/en/latest/calculator_info/sensitivity.html}{online documentation} for details on its implementation). The Compton $y_{\mathrm{t{\textsc{sz}}}}$ sensitivity estimates are instead computed by considering the in-band integral average of the surface brightness-to-Compton $y_{\mathrm{t{\textsc{sz}}}}$ weighted with respect to the atmospheric transmittance levels over the considered frequency range. Here we notice that the actual sensitivity of each individual band to the SZ effect can be significantly hampered by astrophysical background contamination, Galactic foregrounds, and correlated atmospheric noise. As such, the values reported should be taken as a rough estimate of AtLAST's performance in the context of SZ studies. For a more accurate estimation of the expected thermal and kinetic SZ noise levels, we refer to the discussion in Sect.~\ref{sec:test:det}. 
    The width of each band was computed as the frequency range that minimizes the output noise root-mean-square level in the corresponding band for a given integration time (we refer to Sect.~\ref{sec:test:bands} for details).
    Finally, the noise levels reported in the last column are equivalent to the noise RMS that one would measure over a solid angle of $1~\mathrm{arcmin^2}$, and is reported in order to facilitate the direct comparison with wide-field survey instruments.\vspace{6pt}}
    \begin{tabledata}{c^c^c^c^c^c^c^c} 
    \header  band & ref.\ frequency &   bandwidth & band edges &    beam &  \multicolumn{2}{c}{\textbf{sensitivity}} & survey noise \\
    \header   --- &           [GHz] &       [GHz] &      [GHz] & [arcsec] & [$\mathrm{\mu Jy~beam^{-1}~h^{1/2}}$] & [$y_{\mathrm{t\textsc{sz}}}~h^{1/2}$] & [$\mathrm{\mu K_{\textsc{cmb}}-arcmin~h^{1/2}}$]  \\
    \row        2 &            42.0 &          24 &     30--54 & 35.34 &         6.60 & $7.43\times10^{-7}$ &  2.40 \\
    \row        3 &            91.5 &          51 &    66--117 & 16.22 &         6.46 & $1.05\times10^{-6}$ &  1.27 \\
    \row        4 &           151.0 &          62 &   120--182 &  9.83 &         7.14 & $2.82\times10^{-6}$ &  1.21 \\
    \row        5 &           217.5 &          69 &   183--252 &  6.82 &         9.22 & $2.32\times10^{-4}$ &  1.86 \\
    \row        6 &           288.5 &          73 &   252--325 &  5.14 &        11.91 & $1.37\times10^{-5}$ &  3.71 \\
    \row        7 &           350.0 &          50 &   325--375 &  4.24 &        23.59 & $2.76\times10^{-5}$ & 12.26 \\
    \row        8 &           403.0 &          38 &   384--422 &  3.68 &        39.98 & $6.31\times10^{-5}$ & 34.70 \\
    \row        9 &           654.0 &         118 &   595--713 &  2.27 &        98.86 & $2.16\times10^{-3}$ & $1.67\times10^3$ \\
    \row       10 &           845.5 &         119 &   786--905 &  1.76 &       162.51 & $3.94\times10^{-2}$ & $3.70\times10^4$ \\
 \end{tabledata}
    \label{tab:freq_sens_beam}
\end{sidewaystable*}

\subsection{Sensitivity estimates and detector requirements}\label{sec:test:det}
As broadly discussed in the previous sections, performing a clean and robust separation of the multiple spectral components comprising the millimeter/submillimeter astronomical sky will present the biggest observational challenge to achieving our proposed SZ science goals. Any contributions from instrumental noise, Galactic foregrounds, and extragalactic backgrounds that are not properly accounted for in the component separation or spectral analyses can  result in residual noise or systematic biases. As such, this represents a limiting factor in the detectability of any SZ signal and could be interpreted as the final SZ depth of the proposed observations.

\subsubsection{Wide-field survey strategy}\label{sec:test:det:wide}
To obtain a straightforward estimate of the instrumental performance expected when adopting the proposed spectral setup (Table~\ref{tab:freq_sens_beam}), we extend the analysis performed by \citet{Raghunathan2022b} and \citet{Raghunathan2023} to simulate an AtLAST-like facility. We refer to these works for a more extended description of the technical details on the simulation products.

In brief, we generate multi-frequency mock sky maps, including contributions from the CMB, radio galaxies, and dusty star-forming galaxies, along with either the kinetic and thermal SZ fields depending on whether we are estimating the thermal or kinetic SZ residuals, respectively. The CMB signal is generated as a random realization with its spectrum given by the lensed CMB power spectrum for our reference cosmology as extracted from \texttt{CAMB} \citep{Lewis2000}. The radio and infrared galaxies are modelled assuming a Gaussian approximation to their Poisson distribution, where we adopt the measurements by the South Pole Telescope \citep{George2015} as reference power spectra. The thermal SZ distribution is obtained by generating a Poisson distribution of haloes according to the \citet{Tinker2008} halo mass function, and then pasting a thermal SZ signal on each halo. 
The thermal SZ signal is modeled using a generalized Navarro-Frenk-White \citep{Nagai2007} which has been calibrated using X-ray observations \citep{Arnaud2010}. 
The dimensionless pressure profile is integrated along the line-of-sight to obtain the Compton $y_{\mathrm{t\textsc{sz}}}$ which is further integrated within the angular extent of the cluster $r_{500}$ to get the integrated cluster Compton $Y_{SZ}$. 
We use the generalization of the Planck $Y_{SZ}$-to-mass scaling relation \citep{Planck2015XXIV} proposed by \citet{Louis2017} and \citet{Madhavacheril2017} to introduce a mass and redshift evolution of the signal. For the kinetic SZ term, we consider a flat power spectrum with amplitude $3~\mathrm{\mu K^2}$, consistent with \citet{Raghunathan2023} and based on the SPT results reported by \citet{Reichardt2021}. Finally, the noise spectrum is assumed to be described as a combination of a white noise component ($\Delta_{T}$) and an atmospheric component as
$N_\ell =  \Delta_{T}^2 \left[1 + \left(\ell/\ell_{\mathrm{knee}}\right)^{\alpha_{\rm knee}}\right]$. 
The white noise levels are given in Table~\ref{tab:freq_sens_beam} and the large-scale atmospheric component is modeled using the $\ell_{\mathrm{knee}}$ and $\alpha_{\mathrm{knee}}$ with values coming from \citet{Raghunathan2022a} and \citet{Raghunathan2022b} assuming a Chilean survey.
For simplicity, we neglect any relativistic and non-thermal SZ component or any Galactic contributions, but note that these could induce major uncertainties in the reconstruction of the thermal and kinetic SZ effects on large scales \citep{Raghunathan2022a}. Similarly, we are here considering the thermal SZ effect from the individual haleos not to be correlated with the signal from radio and dusty star-forming galaxies. However, for clusters close to the SZ confusion noise or in the presence of significant star formation and radio activity (e.g., in the case of protocluster complexes; Sect.~\ref{sec:theory:proto}), the spatially correlated contamination from (sub)millimeter bright sources could degrade the overall sensitivity to the thermal SZ signal. Still, \citet{Raghunathan2022b} showed that, in the case of wide-field CMB surveys like CMB-S4, the inclusion of a correlated component in the contamination from compact sources would bias the thermal SZ estimtes low by $\lesssim0.2\sigma$. In the case of AtLAST, we expect that the possibility of exploiting its high angular resolution (and, thus, the lower confusion noise) for enhancing the ability of performing a spatial component separation will allow for minimizing the bias from any spatially correlated contamination. We aim a properly testing this in a future work, and, for the moment, we limit ourselves to referring to the analysis in the aforementioned work by \citet{Raghunathan2022b}. We further refer to this for a detailed discussion of the effect of thermal SZ confusion noise in the context of wide-field SZ surveys, along with a strategy for mitigating its impact on the detection of low-mass haloes.
 
\begin{figure*}[!ht]
    \centering
    \includegraphics[clip,trim=0 0 0 0,width=0.98\textwidth]{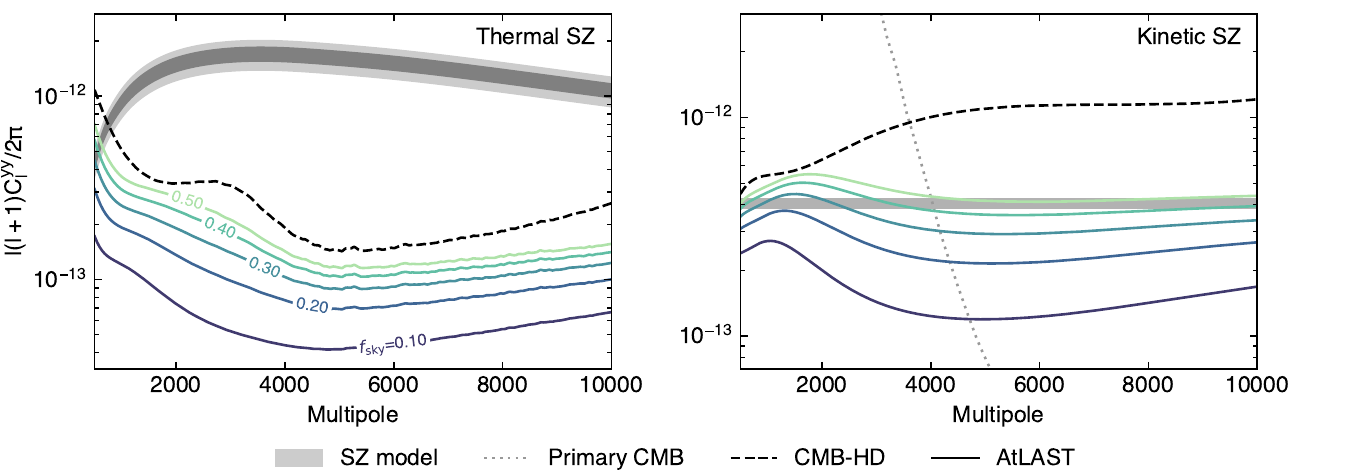}
    \caption{Noise power spectra for the thermal SZ Compton $y_{\mathrm{t\textsc{sz}}}$ (left panel) and kinetic SZ Comptony $y_{\mathrm{k\textsc{sz}}}$ (right panel) residuals as a function of the sky coverage ($f_{\mathrm{sky}}$) in the case of a wide-field AtLAST survey (adapted from \citealt{Raghunathan2022a} and \citealt{Raghunathan2023}, which we refer to for details). In both panels, we include as a reference the residual noise curve expected in the case of the CMB-HD survey (dashed black line; \citealt{CMBHD2019}). The dotted gray line in the right panel shows the temperature power spectrum for the primary CMB converted to Compton $y_{\mathrm{k\textsc{sz}}}$ units following Eq.~\ref{eq:ksz}. The shaded band in the left panel denotes the power spectrum for our fiducial thermal SZ sky based on \citet{Battaglia2010} with the signal level and the $1\sigma$ and $2\sigma$ credible intervals scaled to match the measurements from Atacama Cosmology Telescope (ACT; \citealt{Choi2020}) and South Pole Telescope (SPT; \citealt{Reichardt2021}). 
    As a reference kinetic SZ signal (gray band in the right panel), we consider the same power spectrum level $\ell(\ell+1)C_{\ell}^{\mathrm{TT}}/2\pi=\ell(\ell+1)C_{\ell}^{\mathrm{yy}}/2\pi ~T_{\textsc{cmb}}^2=3~\mathrm{\mu K^2}$ adopted by \citet{Raghunathan2023} based on the results from SPT \citep{Reichardt2021}. We note that extending the model to multipoles $l\gtrsim10\,000$ would necessarily require a major extrapolation of the thermal and kinetic SZ spectral models over angular scales falling below the resolution of state-of-the-art CMB survey experiments and for which we currently have no constraints. AtLAST will be uniquely positioned to probe at high significance the thermal and kinetic SZ power spectra over this unexplored range of angular scales.}
    \label{fig:survey}
\end{figure*}

In Figure~\ref{fig:survey} we present the result of an optimal internal linear combination of the simulated multi-frequency AtLAST maps when adopting a wide-field survey strategy over a period of 5 years and when covering different sky fractions. AtLAST will be able to probe both the thermal and kinetic SZ signals down to Compton $y_{\mathrm{t\textsc{sz}}}$ and $y_{\mathrm{k{\textsc{sz}}}}$ levels $\lesssim 5\times10^{-7}$ over an extreme dynamic range of spatial scales when considering a deep blind survey approach ($<4000~\mathrm{deg^2}$). As shown previously in Fig.~\ref{fig:masszeta}, this implies that the proposed spectral configuration will allow AtLAST to reach, at better angular resolution, a mass limit almost $2\times$ lower than CMB-HD \citep{CMBHD2019}, a reference concept for a next-generation 30-meter CMB experiment.

\subsubsection{Targeted observations}
When considering narrow-field targeted observations, we estimate that AtLAST will be able to reach a beam-level Compton $y$ depth of $\sim2\times10^{-6}$ per hour of integration time. Previous studies \citep[e.g.,][]{Dolag2016,Raghunathan2022b} have predicted that the confusion noise floor due to the thermal SZ signal from low-mass haloes --- either in the proximity of more massive systems due to physical association or projection effects, or falling below a given detection threshold --- amounts to $y_{\mathrm{t\textsc{sz}}}\simeq2-5\times10^{-7}$ in the case of $<10^{13}~\mathrm{M_{\odot}}$ galaxy clusters (corresponding to the predicted detection threshold for AtLAST). As such, the estimated sensitivity implies that AtLAST will obtain SZ-confusion-limited observations in $\sim100~\mathrm{h}$ of on source integration time in deep, targeted maps. Nevertheless, this sensitivity estimate corresponds to the residual Compton $y$ root-mean-square noise obtained when applying a constrained internal linear combination procedure to a simple set of mock targeted observations. We would like to stress that the generation of these simulated AtLAST maps and the consequent sensitivity estimation differs from the analysis presented above for a  wide-field survey case. In particular, we generate flat-sky sky realizations at the AtLAST bands including, differently than \citet{Raghunathan2022b}, both Galactic foregrounds and extragalactic background components. This is intended to propagate any uncertainties associated with the de-projection of large-scale contamination and with the Poisson distribution of extragalactic sources. In particular, the foreground model is based on the \texttt{pysm3} models \citep{Thorne2017}, but we ported and adapted the code to extend the stochastic components down to arcsecond scales. The output mock realization comprises the dust (\texttt{d11} model), AME (\texttt{default}), free-free (\texttt{default}) and synchrotron (\texttt{s6}) from the Milky Way. Further, we consider a background signal comprising a random CMB realization, as well as infrared and radio backgrounds from unresolved sources as extracted from the SIDES \citep{Bethermin2017} and RadioWebSky simulations \citep{Li2022}, respectively. To reproduce a clean subtraction of any dominant contaminating compact sources, we excluded all the radio and infrared components with fluxes in at least two bands larger than $3\times$ the corresponding noise root-mean-square. As such, it should be considered as a rough ground reference for the actual depth achievable with future AtLAST measurements. Further, we currently lack accurate models for any of the aforementioned model components down to arcsecond scales, and the adopted extrapolation could introduce unconstrained biases in the noise estimates. Future forecasting studies will particularly investigate how different observation strategies, source subtraction, and modeling techniques will affect the contamination mitigation and the effective SZ sensitivity.

\subsubsection{Optimal detector count}
\begin{figure}[!ht]
    \centering
    \includegraphics[clip,trim=0 0 0 10pt,width=0.49\textwidth]{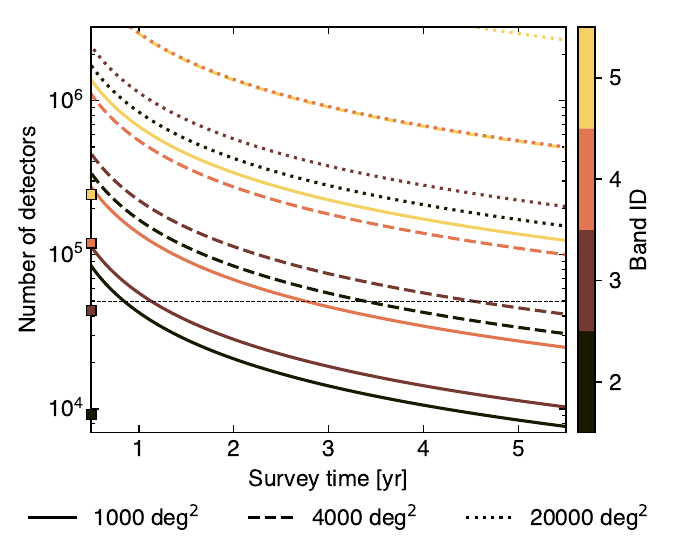}
    \caption{Required number of detectors to reach the target sensitivity estimates listed in Table~\ref{tab:freq_sens_beam} for different bands and considering different survey strategies. For comparison, the squares on the ordinate marks the detector counts required in each band for fully covering a $1~\mathrm{deg^2}$ field of view. The horizontal line traces the minimal number of detectors ($50\,000$) identified for reaching the target depth in Band 2 and 3 in the case of $1\,000~\mathrm{deg^2}$ and $4\,000~\mathrm{deg^2}$ surveys. We note that this is consistent with the estimated specifications reported in the \href{https://www.atlast.uio.no/memo-series/memo-public/instrumentationwgmemo4_29feb2024.pdf}{AtLAST Memo 4} for the 1\textsuperscript{st} generation instruments.}
    \label{fig:survey_ndet}
\end{figure}

Achieving the frontier capabilities discussed in the previous sections will necessarily demand a considerable mapping speed and, thus, a crucial effort in the optimization of the detector array. To estimate a minimal detector count for filling the focal plane, we consider the sensitivity estimates reported in Table~\ref{tab:freq_sens_beam} as target depths for surveys with varying observing period and sky coverage (see Figure~\ref{fig:masszeta}). The results are reported in Figure~\ref{fig:survey_ndet}. A detector count $n_{\mathrm{det}}\simeq5\times10^4$ is sufficient for achieving the sensitivity goal in the case of a narrow survey configuration ($1000~\mathrm{deg^2}$) both in Band~2 and Band~3, key for tracing the decrement regime of the thermal SZ signal. For the same bands, the same $n_{\mathrm{det}}$ constraints would allow us to achieve a similar survey sensitivity also in the intermediate $4000~\mathrm{deg^2}$ case over $\sim4-5~\mathrm{years}$. Nevertheless, extending these considerations to other bands or a wide-field scenario would require a significant increase in $n_{\mathrm{det}}$. For instance, in the case of Band~5 --- crucial for constraining the departures from the thermal SZ effect due to kinetic and relativistic contributions --- such a boost would range over almost an order of magnitude. 

As broadly highlighted in Sect.~\ref{sec:theory}, constraining the small-scale fluctuations in the thermal and kinetic SZ effects, while tracing the temperature-dependent relativistic SZ corrections would imply measuring deviations from the global SZ distribution order of magnitudes smaller than the bulk, non-relativistic thermal SZ signal. This would in turn require a significant reduction of any systematic effects hampering the overall calibration accuracy.
In this regard, an interesting technical aspect of AtLAST is the plan for integrating a continuous control system feedback loop relying on monitoring sensors to track in real time any deformations in the primary mirror alignment in order to provide rapid response compensation or corrections. Such a ``closed-loop'' control system will move beyond more common ``open-loop'' solutions based on the estimation of deformation corrections from external information (e.g., pre-computed look-up tables, or dedicated calibration observations for astro-holography), and will allow for actively tracking the alignment of the primary mirror panels (we refer to \citealt{Mroczkowski2025} and \citealt{Reichert2024} for details). Ongoing developments --- such as the laser system for the SRT \citep{Attoli2023} or the wavefront sensing system on the NRO 45-m telescope \citep{Tamura2020,Nakano2022} -- are now showing that active closed-loop metrology systems can keep the errors in the beam down to sub-percent levels, meaning the beam will be diffraction limited and stable throughout observations. This in turn will improve the calibration accuracy and reduce systematics (see, e.g., \citealt{Naess2020} for discussion of the diurnal effects on the ACT beams) that have been shown to impact CMB and SZ results at the several percent level in the case of passive optics ($3-5\%$; \citealt{Hasselfield2013,Lungu2022}), with the result that the daytime data have generally been excluded from cosmological analyses. Future dedicated forecasts will analyze the benefits of metrology for secondary CMB measurements using AtLAST, including improvements to the calibration, reduction of systematics, and the ability to recover larger angular scales on sky. However, the salient takeaway message is that uncertainties in the beam should no longer be a leading source of systematic error.

\subsection{Mock reconstruction of the relativistic SZ effect}\label{sec:test:rtsz}
The relative amplitude of the relativistic component compared to the thermal and kinetic SZ effects makes this modeling task highly challenging. To test the prospects of using AtLAST measurements for performing a spectral separation and analysis of the SZ effect, we thus perform a mock reconstruction of the intracluster temperature using the relativistic SZ effect.

\subsubsection{SZ-only reconstruction}\label{sec:test:rtsz:case1}
\begin{figure}[!ht]
    \centering
    \includegraphics[width=0.49\textwidth]{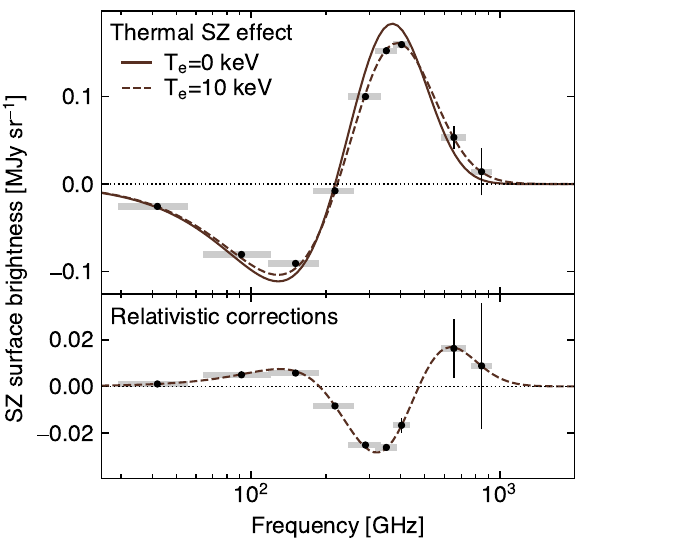}
    \caption{
    Predicted rtSZ measurements for a cluster with a temperature of $10~\mathrm{keV}$ and Compton $y_{\mathrm{t\textsc{sz}}}$ of $10^{-4}$, assuming flat bandpasses in the Bands 2--10 (denoted as gray bands; see also Table~\ref{tab:freq_sens_beam}). For a comparison, we include the spectral signature for the classical formulation of the the thermal SZ effect, i.e., derived under the non-relativistic assumption for the ICM electrons ($T_{e}=0~\mathrm{keV}$; see Sect.~\ref{sec:sz:tsz} and \ref{sec:sz:rsz} for a discussion). We assume the same exposure time in each band and account for flux sensitivity and beam size differences, tuned to achieve a $\mathrm{S/N}=50$ in Band 8. For comparison, the non-relativistic approximation is also shown. The bottom axis shows the difference between the relativistic and non-relativistic spectra.}
    \label{fig:rtSZ_spec}
\end{figure}

As a test case, we consider a galaxy cluster with temperature $T_{\mathrm{e}} = 10~\mathrm{keV}$ and Compton $y_{\mathrm{t\textsc{sz}}}= 10^{-4}$. We note that, despite representing relatively extreme (but realistic) values, the setup $(T_{\mathrm{e}},y_{\mathrm{t\textsc{sz}}})=(10~\mathrm{keV},10^{-4})$ is chosen to facilitate this first study of the AtLAST capabilities of providing spectral constraints on temperature-dependent distortions of the thermal SZ effect. A broader exploration of the parameter space will be presented in Sect.~\ref{sec:test:rtsz:case3}. The amplitude of the SZ signal at each of the selected bands in the minimal spectral set is obtained by integrating the relativistically-corrected thermal SZ (rtSZ) spectrum across each band assuming flat bandpasses. We then obtained estimates for the corresponding uncertainties based on the sensitivity estimates from the AtLAST sensitivity calculator \citep{senscalc}. First, we compute the integration time required to achieve S/N $\geq$ 50 in Band 8, arbitrarily chosen among the two spectral windows closest to the peak in the rtSZ effect (Figures~\ref{fig:sz_spectrum} and \ref{fig:rtSZ_spec}). The resulting noise root-mean-square is defined as the corresponding uncertainty. The uncertainties $\delta I$ for each of the remaining bands are thus computed assuming the same integration time as for the Band 8 estimation above, but taking into account both the differing point-source sensitivities and beam sizes across frequency bands,
\begin{equation}
\frac{\delta I(n)}{\delta I(\mathrm{8})} = \frac{\sigma(n)}{\sigma(\mathrm{8})} \times \frac{\Omega(\mathrm{8})}{\Omega(n)}.
\end{equation}
Here, $n$ denotes the band index ($n=\{2,...,10\}$), while $\sigma$ and $\Omega(\nu)$ are the flux density RMS and the beam size at a given frequency $\nu$, respectively. The resulting simulated measurements are shown in Fig.~\ref{fig:rtSZ_spec}.

\begin{figure}[tbh]
    \centering
    \includegraphics[width = 0.49\textwidth]{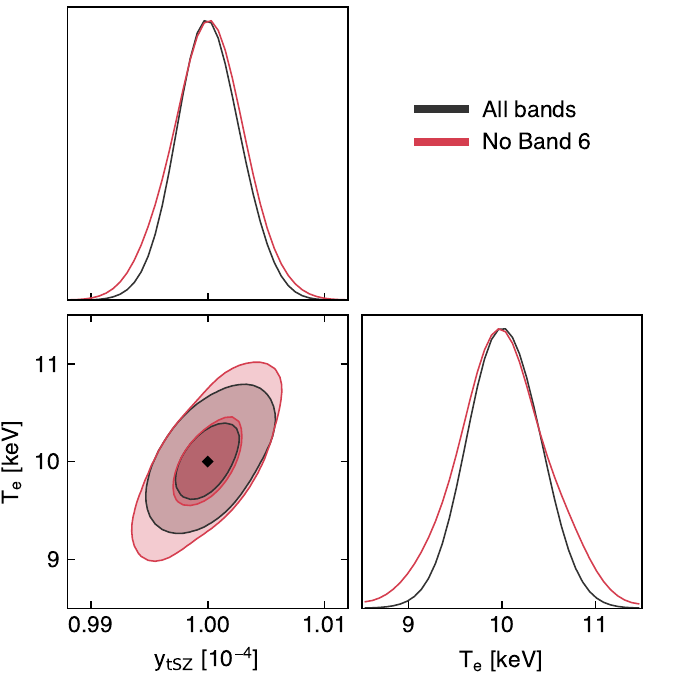}
    \caption{
    Fits to the simulated measurements shown in Figure~\ref{fig:rtSZ_spec}, using all bands (grey) and all except Band 6 (red). The black diamond denotes the input parameters. Despite these are recovered accurately, excluding Band 6 increases the $y_{\mathrm{t\textsc{sz}}}$-$T_{\mathrm{e}}$ degeneracy substantially.}
    \label{fig:rtSZ_temp_fit}
\end{figure}

The derived SZ measurements can then be used to perform a simple joint inference of the Compton $y_{\mathrm{t\textsc{sz}}}$ and electron temperature for the target case. If only lower-frequency data points ($\lesssim200~\mathrm{GHz}$) are measured, then there is a complete degeneracy between $T_{\mathrm{e}}$ and the Compton $y_{\mathrm{t\textsc{sz}}}$ parameter. In this spectral range, in fact, an increase in the electron temperature reduces the signal in a similar manner as decreasing the overall Compton $y_{\mathrm{t\textsc{sz}}}$ amplitude. When higher-frequency points are included, the degeneracy can be minimized as shown in Figure~\ref{fig:rtSZ_temp_fit}. By dropping one band at a time from the fit, we find that Band 6 ($\approx$\,240\,GHz) has the greatest influence in breaking the degeneracy since the S/N on the difference between nearly-degenerate models is greatest in this band.

\subsubsection{SZ+dust reconstruction}\label{sec:test:rtsz:case2}
\begin{figure}[tbh]
    \centering
    \includegraphics[width=0.49\textwidth]{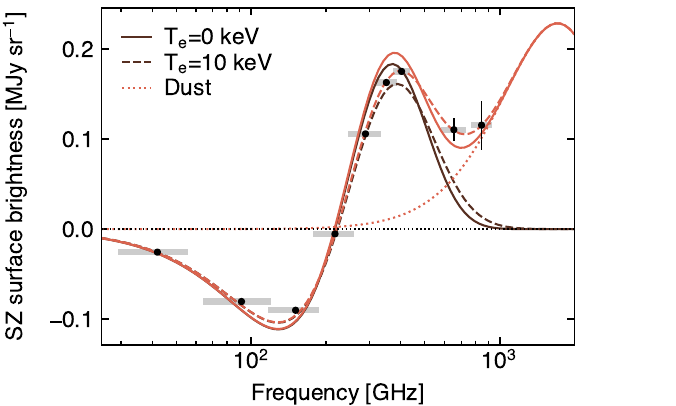}
    \caption{Same as Figure~\ref{fig:rtSZ_spec}, but with the addition of a modified black body dust component based on the model from \citet{Erler2018}. The red dotted line shows the dust signal. The red solid and dashed lines show the total signal from the dust and non-relativistic and relativistic signals respectively.}
    \label{fig:rtSZ_dust_spec}
\end{figure}

So far, we have assumed that the only signal present is the SZ signal. However, in reality there will of course be other astrophysical foregrounds and backgrounds present along the line of sight, resulting in non-negligible contamination of the overall SZ signal observed in the direction of a galaxy cluster. In this test, we however assume that signals that are not spatially correlated to the SZ effect can be removed by means of component separation methods (see Sect.~\ref{sec:test:det}) or targeted forward modeling procedures in the case of unresolved compact sources \citep[e.g.,][]{Ruppin2017,DiMascolo2019,Keruzore2020,Kitayama2020,Andreon2021,Paliwal2024}. However, previous studies (e.g. \citealt{Erler2018}) showed that there is a spatially correlated signal within clusters associated with the diffuse dust emission. To understand the impact on the capability of AtLAST in constraining any rtSZ deviation, we repeat the above test by adding an additional dust-like spectral component (Figure~\ref{fig:rtSZ_dust_spec}). In particular, we assume a modified black body signal given by \citep{Erler2018}

\begin{equation}
    I_{\mathrm{dust}}(\nu) = A_{\mathrm{dust}}^{857} \left ( \frac{\nu}{\nu_0} \right )^{\beta_{\mathrm{dust}}+3} \frac{\exp \left [ h \nu_0 / (k_B T_{\mathrm{dust}}) \right ] -1}{\exp \left [ h \nu / (k_B T_{\mathrm{dust}}) \right ] -1},
\end{equation}

where $\nu_0=857$\,GHz is chosen as the reference frequency, and $A_{\mathrm{dust}}^{857}$ is the amplitude at this frequency. To generate a dust signal, we use the parameter fits for $A_{\mathrm{dust}}^{857}$, $\beta_{\mathrm{dust}}$ and $T_{\mathrm{dust}}$ as provided by \citet{Erler2018}, and add them as free parameters to our fit with uniform priors on all parameters. The uncertainties on the measurements in each band are the same as in the SZ-only fit (Sect.~\ref{sec:test:rtsz:case1})

\begin{figure*}[tbh]
    \centering
    \includegraphics[width=0.80\textwidth]{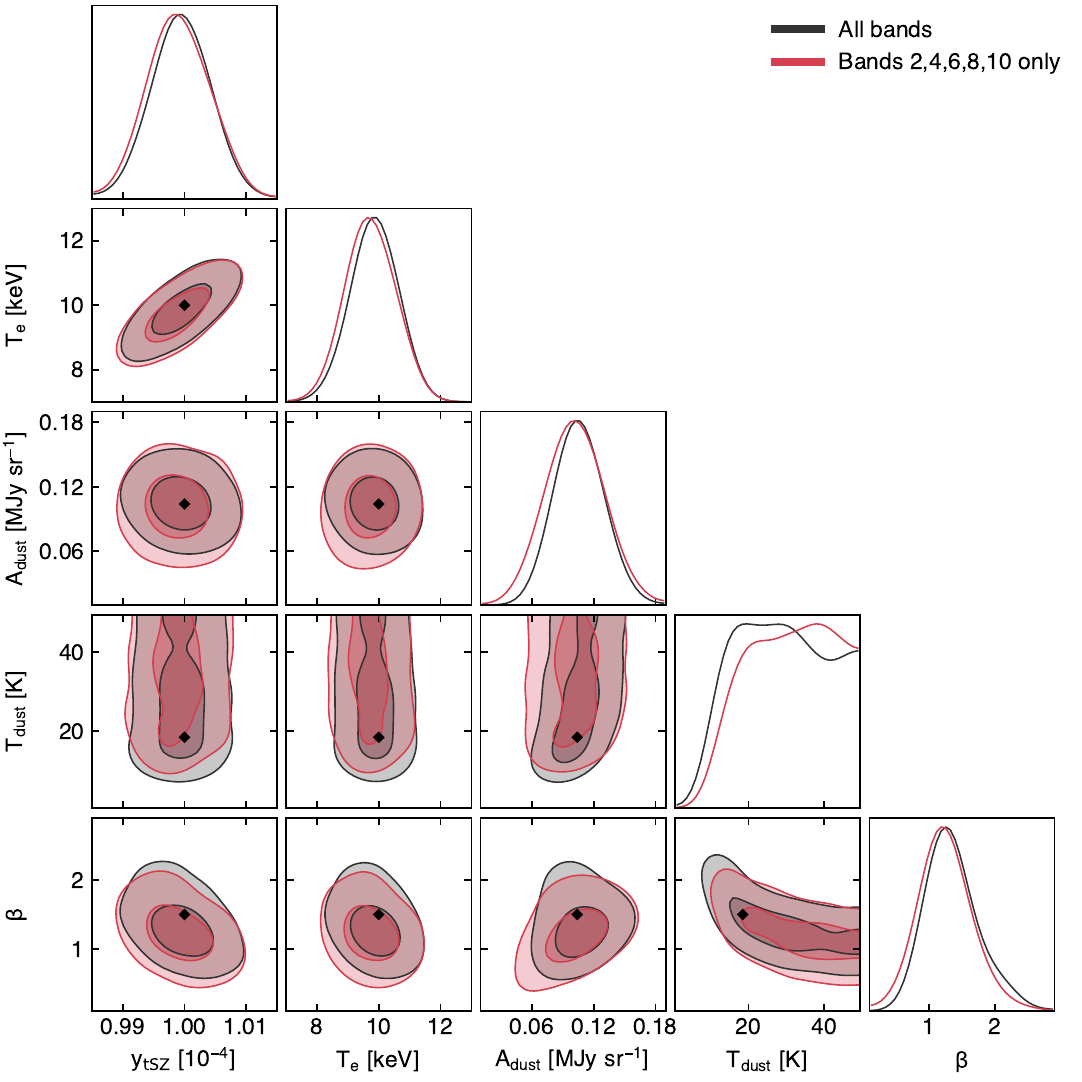}
    \caption{
    Fits to the simulated measurements shown in Figure~\ref{fig:rtSZ_dust_spec}, using all bands (grey) and Bands 2, 4, 6, 8 and 10 only (blue). With this optimal set of five bands, the constraints on the rtSZ parameters are almost equivalent to the constraints with all bands, but we obtain a slightly reduced constraining power on the dust parameters. The dotted diamonds denote the input parameters.}
    \label{fig:rtSZ_dust_fit}
\end{figure*}

In this case, more bands become necessary to correctly constrain the rtSZ temperature and disentangle the rtSZ and dust spectral components. The best minimal combination comprises Bands 2, 4, 6, 8 and 10, that provide almost identical constraints to the full set of bands on the rtSZ parameters, while achieving a lower precision on the dust parameters (as shown in Figure~\ref{fig:rtSZ_dust_fit}). Most importantly, it is important to note that the broad spectral coverage offered by the proposed setup allows a clean separation of the rtSZ and dust signals with only a marginal impact on the rtSZ constraints compared to the SZ-only case (Sect.~\ref{sec:test:rtsz:case1}). 

\subsubsection{Required sensitivity and time forecasts}\label{sec:test:rtsz:case3}
The reference S/N of 50 employed above was mainly intended to achieve a general perspective on the spectral constraining power of the proposed setup without being limited by the inherent significance of the test SZ signal. Thus, we now investigate what S/N is required to achieve good temperature constraints from rtSZ measurements. In particular, we run similar fits for different values of the reference S/N and different temperatures.

\begin{figure}[tbh]
    \centering
    \includegraphics[width=0.49\textwidth]{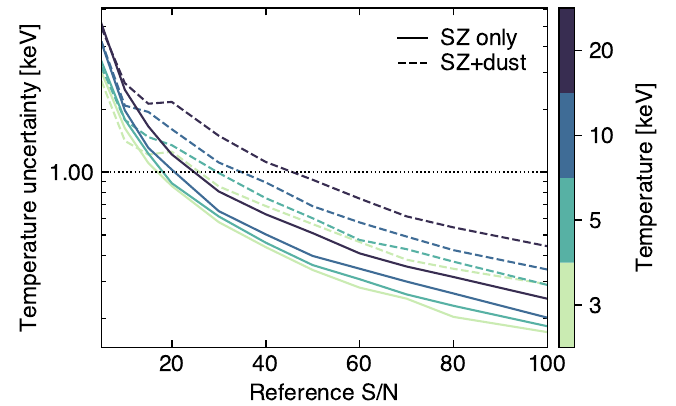}
    \caption{Precision achieved on the temperature reconstruction as a function of the signal-to-noise (S/N) in the reference band ($384-422~\mathrm{GHz}$; Band 8) and for a range of ICM temperatures. The horizontal line denotes the target temperature accuracy $T_{\mathrm{e}}=1~\mathrm{keV}$ discussed in the text. A level of $\mathrm{S/N}\gtrsim40$ will be sufficient for constraining the temperature in any of the tested cases, also in the presence of dust contamination.}
    \label{Fi:rtSZ_temp_SNR}
\end{figure}

The impact of the varying S/N on the temperature reconstruction is summarized in Figure~\ref{Fi:rtSZ_temp_SNR}. If we require, for example, an unbiased reconstruction of the electron temperature with a precision of $1~\mathrm{keV}$, a reference $\mathrm{S/N}\simeq30$ is sufficient for all the temperatures tested. When dust is included, S/N should increase to levels $\sim40$ to be able to constrain the temperature for all except the hottest clusters at the reference precision. Such an effect is caused by the increased impact of relativistic corrections on the high frequency end of the SZ spectrum for the hotter clusters, in turn introducing a more prominent degeneracy with the dust spectral component.

Although AtLAST will be able to observe clusters spanning a broad range in mass and redshift (and, hence, temperature), the example analysis presented in the previous section is aimed only at forecasting AtLAST capabilities of measuring relativistic deviations from the standard thermal SZ and not at testing the expected detection threshold as a function of cluster properties. To take into account the evolution of the rtSZ effect with the mass and redshift of a galaxy cluster, we aim here at estimating the required observing time to reach a target S/N in Band 8, our reference spectral window (Sect.~\ref{sec:test:rtsz:case1}).

\begin{figure}[!ht]
    \centering
    \includegraphics[clip,trim=0 5pt 0 0,width=0.49\textwidth]{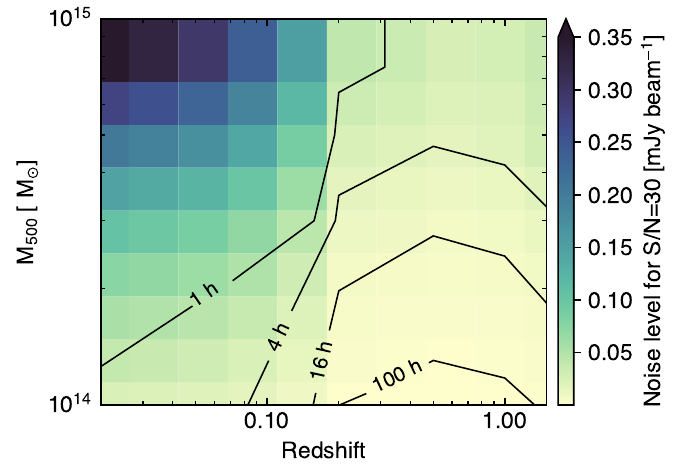}
    \caption{
    Beam-level noise root-mean-square and observing times as a function of cluster redshift and mass $M_{200}$ required to reach an S/N of 30 in the reference spectral band ($384-422~\mathrm{GHz}$; Band 8), allowing average SZ temperatures to be well constrained. AtLAST will enable inexpensive measurements of the ICM temperature for the most massive systems in the local universe ($z\lesssim0.2$), but the expected integration time remains relatively limited also for the less massive and distant systems.}
    \label{fig:rtSZ_global_times}
\end{figure}

To do so, we construct cluster signal maps for a range of masses and redshifts, using the physical model given in \citet{Olamaie2012} and \citet{Javid2019}.  Assuming hydrostatic equilibrium, this model gives us physically consistent pressure and temperature profiles which we use with the \texttt{SZpack} (\citealt{Chluba2012}, \citealt{Chluba2013}) temperature-moment method to predict relativistic SZ effect signal maps, taking into account the spatial variation of the temperature. The resulting observing time predictions for a reference S/N of 30 are shown in Figure~\ref{fig:rtSZ_global_times}. For reasonable observing times ($<16~\mathrm{hours}$ on source) we can get average temperature constraints for most clusters at redshifts up to $z\approx0.1$, and high-mass clusters ($M_{200}\gtrsim 4\times 10^{14} \mathrm{M}_{\odot}$) up to arbitrarily high redshift. It is important to note that the enhanced angular resolution of AtLAST could easily allow one to obtain spatially resolved information on the temperature distribution, once the S/N requirements are satisfied for each spatial element considered for the analysis --- e.g., radial bins or spectrally homogeneous regions as generally considered in high-resolution X-ray studies (e.g., \citealt{Sanders2006}).

\section{AtLAST SZ studies in a multi-probe context}\label{sec:synergy}

AtLAST will provide an unprecedented speed and spectral grasp across the (sub)millimeter spectrum. This will make AtLAST inherently relevant beyond just SZ science, and will open up possibilities for fundamental synergies in the multi-wavelength and multi-probe exploration of the Universe.

\subsection{AtLAST scientific cross-synergies}
Thanks to the novel multi-instrument design \citep{Mroczkowski2025}, AtLAST will be aimed at representing a high-impact (sub)millimeter facility with a broad and varied scientific reach \citep{Booth2024a,Booth2024b}. As such, this will set the ground for a natural cross-synergy across the different scientific applications identified as part of the AtLAST Science Development effort. 

In the case of a wide-field continuum survey discussed in Sect.~\ref{sec:test:det}, the multi-band coverage and the extended temporal span will make the SZ-driven observations extremely valuable for temporally-dependent studies as for transient surveys \citep{OrlowskiScherer2025}.
Similarly, the multiple bands and likely polarization sensitivity will be useful in the study of Galactic dust, molecular clouds, and circumstellar discs \citep{Klaassen2024}, building on the lower resolution results with, for example, the Simons Observatory \citep{Hensley2022}.

Related to the science goals proposed in this work, the availability of a wide-field spectroscopic survey of the distant Universe \citep{vanKampen2024} will immediately enhance the validity of the SZ identification and study of high-$z$ clusters and protoclusters by providing accurate redshift information. The broad spectral coverage achieved thanks to the proposed multi-band setup will actually play a crucial role in maximizing the redshift domain. At the same time, as already mentioned in previous sections, having access simultaneously to constraints on the physical properties of large-scale environments via the SZ effect (Sect.~\ref{sec:theory:proto}) and on the associated galaxy populations (via the spectral characterization of their cold molecular gas and the inference of their dust content; see Figure~\ref{fig:dusty}) will represent an unprecedented opportunity in the context of galaxy-environment co-evolution studies. 

Similarly, the information on the warm/hot component of galactic haloes will be essential for building a comprehensive picture of the diffuse and multi-phase CGM \citep{Lee2024}. A combination of the novel perspective offered by AtLAST on the cold contribution with the tight measurements of the thermal and kinetic properties of such elusive haloes (Sect.~\ref{sec:theory:AGNfeedback}) will represent the only way for shedding light on the many potential evolutionary routes of the elusive large-scale CGM.

\subsection{Synergies with other state-of-the-art and forthcoming facilities}\label{sec:synergy:other}
In the context of large scale structures, AtLAST's constraints on the multi-faceted SZ effect will be highly complementary to multi-wavelength information on the galaxy motions and distribution, the gravitational potentials of the systems, the X-ray emission, magnetic field structure, and the highly non-thermal and relativistic emission traced by radio emission. Below, we highlight some of the key facilities and experiments that provide the most synergy with AtLAST.

\subsubsection{Radio}\label{sec:synergy:radio}
The radio waveband offers information that can complement and enhance many of the science cases outlined above, and next-generation instruments such as the Square Kilometer Array (SKA; \citealt{Huynh013}) and the next-generation Very Large Array (ngVLA; \citealt{Selina2018}) will have the angular resolution and sensitivity required to provide it. On the one hand, getting a clear view of the resolved SZ effect and searching for intrinsic scatter and surface brightness fluctuations requires sensitive detection and removal of contaminating radio sources (e.g., \citealt{Dicker2021,Dicker2024}).  While this will already be possible with AtLAST's data itself thanks to its spectral coverage and $\approx 5\arcsec \lambda~\rm mm^{-1}$  resolution (i.e. 10\arcsec\ at 2 mm), interferometric observations at lower frequency (SKA-MID) will aid in pinpointing the location and morphology of the radio sources, while also being more sensitive to fainter sources as most will be brighter at lower frequency.

On the other hand, radio information provides a powerful complementary probe of the astrophysics in the cluster, being sensitive to magnetic fields and populations of non-thermal electrons. The reference surveys proposed for the SKA \citep{Prandoni2015} predict the detection of $\sim1\,000$s of radio haloes with SKA1-LOW, out to redshifts of at least $0.6$ and masses $M_{500}>10^{14}~\mathrm{M_{\odot}}$ (\citealt{Cassano2015,Ferrari2015}; see also, e.g., \citealt{Knowles2021,Knowles2022,Duchesne2024} for preliminary results from SKA precursors), along with potential first detections of the polarization of radio haloes \citep{Govoni2015}. This offers the opportunity not only to compare the detailed astrophysics of the thermal and non-thermal components of clusters, shedding light on the turbulent properties currently limiting the accuracy of mass estimation (see Section~\ref{sec:theory:profile}), but also to potentially discover new populations of clusters via their radio signals. When also observed by AtLAST, these populations will offer insight into the variation in cluster properties when selecting by different methods (see also Section~\ref{sec:theory:elusive}). Faraday Rotation Measure observations of polarized sources behind galaxy clusters as well as studies of tailed radio galaxies \emph{in} clusters will enable the study of cluster magnetic fields in unprecedented detail \citep{Bonafede2015,JohnstonHollit2015a,JohnstonHollit2015b}, contributing to our astrophysical understanding and ability to make the realistic simulations crucial for interpreting observations. 

Ultimately, the direct correlation of the SZ information with the spatial, spectral, and polarimetric properties of the multitude of radio structures observed in the direction of galaxy clusters will be essential for constraining the detailed mechanisms governing particle (re)acceleration within the ICM \citep{vanWeeren2019}. 
Specifically, there is mounting evidence that the non-thermal plasma observed in the form of (multi-scale) radio haloes \citep[e.g.,][]{Gitti2015,Cuciti2022} as well as intercluster bridges \citep[e.g.,][]{Botteon2020b,Bonafede2022,Radiconi2022} originates due to turbulent (re)acceleration \citep{Brunetti2007,Brunetti2014,Eckert2017,Cassano2023}. On the other hand, radio relics are connected to (re)acceleration at shock fronts \citep{Akamatsu2013,vanWeeren2017,Botteon2020}. While it is clear that cluster mergers are driving both processes (turbulence and shocks), our understanding of the physics of (re)acceleration in clusters is limited by two factors: (i) information about the distribution of gas motions in the ICM is currently sparse and usually inferred via indirect methods \citep[for a review, see][]{Simionescu2019}, and (ii) characterizing shocks in the low-density cluster outskirts, where radio relics are usually found, is very challenging. 
Detailed mapping of the thermal (sensitive to shocks) and kinetic (sensitive to gas motions) SZ signals throughout the volume of a large sample of galaxy clusters (potentially extending out into the cosmic web), and how these signals relate to features observed in the radio band, will be invaluable towards painting a clear picture of the connection between large-scale structure assembly, magnetic field amplification, and cosmic ray acceleration.

Understanding the impact of AGN feedback, on the other hand, (Section~\ref{sec:theory:AGNfeedback}) requires complementary observations of the AGN themselves. \citet{Gitti2015} finds that even with early SKA1 (50\% sensitivity), all AGN with luminosity $>10^{23}~\mathrm{W~Hz^{-1}}$ can be detected up to $z\leq 1$ with subarcsecond resolution, and the radio lobes thought to be responsible for carving out the X-ray cavities should be detectable in any medium -- large mass cluster at any redshift in the SKA1-MID deep tier surveys.  Moreover, SKA1-MID is predicted to detect intercluster filaments at around $2.5$ -- $6\sigma$ \citep{Giovannini2015}, providing information on their magnetic fields as a complement to the SZ information on their thermodynamic properties (Section~\ref{sec:theory:whim}).

At the top of the SKA frequency range, it will be possible to directly access thermal SZ information. Future extensions to the SKA-MID Phase 1 setup (with the integration of the high-frequency Band 6; we refer to the \href{https://www.skao.int/sites/default/files/documents/d38-ScienceCase_band6_Feb2020.pdf}{SKA Memo 20-01} for details) and the ones envisioned for \href{https://www.skao.int/en/science-users/118/ska-telescope-specifications}{SKA Phase 2} (2030+) will allow SKA to probe the low-frequency ($\lesssim 24~\mathrm{GHz}$) domain of the SZ spectrum, less affected by kinetic and relativistic deviations than the range probed by AtLAST. In fact, \citet{Grainge2015} find that 1 hour of integration is sufficient for obtaining a $14\sigma$ detection of the SZ effect from a $M_{200}=4\times 10^{14}~\mathrm{M_{\odot}}$ cluster at $z>1$. On the other hand, the clean perspective offered by AtLAST on the multiple SZ components will provide the means, e.g., for cleanly disentangling the SZ footprint of galaxy clusters from the faint, diffuse signal from mini- to cluster-scale radio haloes, large-scale relics, and back-/foreground and intracluster radio galaxies, enhancing their joint study.

\subsubsection{Millimeter/submillimeter}\label{sec:synergy:mm} 
Millimetric/submillimetric survey experiments like SO \citep{SO2019} and its upgrades, CCAT-prime/FYST \citep{CCAT2023}, upgrades to the South Pole Telescope \citep{SPT3Gp2022}, and ultimately CMB-S4 \citep{CMBS42016} will cover roughly half the sky over the next few years to a decade, predominantly in the Southern sky. Along with past and current facilities, these however have $\sim$\,arcmin resolution, well-matched to the typical angular size of clusters in order to optimize their detection but not optimal for peering inside clusters to explore astrophysical effects (aside from few nearby exceptional clusters). Nevertheless, while limited to resolutions approximately $\sim8\times$ lower than AtLAST at the same frequencies, their data will provide robust constraints at large scales, lending itself naturally to joint map-making and data combination, as well as valuable source finders for deep AtLAST follow-up. On the other hand, AtLAST will be able to resolve any structures probed by these wide-field surveys, defining a natural and intrinsic synergy.

At still higher resolutions, ALMA will undergo a number of upgrades improving its bandwidth and sensitivity over the next decade.  These upgrades are called the Wideband Sensitivity Upgrade (WSU; \citealt{Carpenter2023}), which in the context of SZ science could deliver $2-4\times$ ALMA's current bandwidth. Wide field mapping capabilities are however not part of the key goals for the WSU, and it is unlikely ALMA will ever map more than a few tens of square arcminutes. Nevertheless, the improved sensitivity and bandwidth could allow for exploiting ALMA to complement AtLAST observations with a high spatial resolution view of astrophysics through detailed follow-up studies.
At the same time, such observations will require AtLAST to recover more extended scales (see Section \ref{sec:technical}). The necessity of such a combination will however allow us to fully leverage the synergistic strengths of single-dish and interferometric facilities for gaining an unprecedented view of the hot baryonic content of the Universe, along with its multi-phase counterparts.

\subsubsection{Optical/infrared}\label{sec:optical-synergies}
The \textit{Euclid} mission \citep{Laureijs2011,Scaramella2022} has recently started surveying the optical/infrared sky, and is expected to result in the identification of $\gtrsim10^{5}$ galaxy clusters and protoclusters across the entire cluster era ($0\lesssim z \lesssim 2$; \citealt{Euclid2019}). Complemented with data from the Legacy Survey of Space and Time (LSST) survey by the forthcoming Vera C.\ Rubin Observatory \citep{Ivezic2019}, these will represent a wealth of complementary constraints on the cluster and protocluster populations that will be essential for enhancing the scientific throughput of AtLAST in the context of SZ studies. The characterization of the weak lensing footprint of galaxy clusters and groups jointly with resolved information on the thermodynamics of their ICM will enable a thorough exploration of the many processes biasing our cluster mass estimates (Sect.~\ref{sec:theory:profile}). At the same time, the detailed characterization of the SZ signal from the vast number of weak-lensing selected systems (and, thus, with different selection effects than surveys relying on ICM properties) will allow for studying in detail the origin of the under-luminous clusters (Sect.~\ref{sec:theory:elusive}).
Similarly, AtLAST will provide an unprecedented view on the ICM forming within the wealth of high-$z$ galaxy overdensities that will be identified by \textit{Euclid}/LSST, in turn providing an unbiased means for constraining the physical processes driving the thermalization of protoclusters complexes into the massive clusters we observe at $z\lesssim2$. 

More in general, the access to a rich set of imaging and spectroscopic measurements by wide-field surveys --- \textit{Euclid}, Rubin Observatory, and the next generation \href{https://roman.gsfc.nasa.gov/}{Nancy Grace Roman Space Telescope} \citep{Spergel2015}, SPHEREx \citep{SphereX2014} --- along with deep, targeted observation from high-resolution facilities --- e.g., JWST \citep{Gardner2006}, or the upcoming \href{https://elt.eso.org/}{Extremely Large Telescope} --- will be greatly complemented by the resolved, wide perspective of AtLAST on the SZ Universe. Tracing the faint warm/hot backbone of large-scale structure (Sect.~\ref{sec:theory:whim}), as well as tightly correlating resolved thermodynamic constraints for the large-scale cluster environment with the physical properties of the galaxies embedded within them \citep{Alberts2022,Boselli2022} and the distribution of the more elusive intracluster light \citep{Contini2014}, will be essential to shed light on their complex and dynamical co-evolution.

In addition, to facilitate precision cosmology studies with \textit{Euclid}/LSST, it is imperative to gain a better understanding of the impact of galactic processes on the redistribution of baryons over large scales. Different prescriptions of feedback employed in various cosmological simulations alter the predicted amplitude and scale dependence of the matter power spectra at separations under 10 Mpc on a level that is considerably larger than the statistical uncertainty expected from upcoming cosmology experiments \citep{Chisari2019,vanDaalen2020}. Mapping the gaseous contents in the low-density outskirts of galaxy groups and WHIM filaments through sensitive AtLAST measurements will thus provide invaluable observational priors necessary to model baryonic feedback for survey cosmology. 

\subsubsection{X-ray}\label{sec:synergy:xray}
On the X-ray side, \href{https://chandra.harvard.edu/}{\textit{Chandra}} and \href{https://www.cosmos.esa.int/web/xmm-newton}{\textit{XMM-Newton}}, launched in 1999 with CCDs capable of $0.5-5\arcsec$ spatial resolution, are still providing a reasonable thermodynamic mapping of the brightest regions of the collapsed structures up to redshift $\sim1.2$. The
\href{https://www.mpe.mpg.de/eROSITA}{\textit{eROSITA}} telescope (launched in 2019) has recently delivered the first release of its X-ray all-sky surveys \citep{Merloni2024} and new catalogs of cluster candidates up to redshift $z\simeq1.3$ \citep{Bulbul2024}. Still, the large point spread function ($\sim15\arcsec$) and the limited sensitivity do not allow for resolving the temperature structure of the ICM and any derived quantities (e.g., pressure, entropy, mass) except for nearby and bright galaxy clusters (e.g., \citealt{Iljenkarevic2022,Sanders2022,Whelan2022,Liu2023}).

The $5~\mathrm{eV}$ spectral resolution of the {\it Resolve} microcalorimeter onboard \href{https://xrism.isas.jaxa.jp/en/}{\textit{XRISM}}, launched in September 2023, has just started a systematic investigation of the gas kinematics in hot, X-ray bright galaxy clusters \citep{XRISM2025I,xrism25_a2029,XRISM2025Coma}. However, these studies will be limited by the low ($\sim 1.3\arcmin$) angular resolution, small field of view and effective area, especially at soft X-ray energies. This particularly hinders the study of less massive haloes (galaxy groups and CGM) and the mapping of extended cluster outskirts and WHIM filaments. 
Forthcoming space missions are expected to improve all these performances through the development of next-generation instruments with higher both spectral and spatial resolutions over a wider field of view and with a larger collecting area: \href{https://www.the-athena-x-ray-observatory.eu/en}{\textit{NewAthena}} (expected to be adopted as a L-mission by ESA in 2027 for a launch in 2037) will outperform the current satellites thanks to the larger effective area by an order of magnitude with a spatial resolution better than 10 arcsecs \citep{newathena25}; 
\href{https://blog.umd.edu/axis/}{\textit{AXIS}} (a Probe mission selected by NASA for Phase A Study) will extend and enhance the science of sensitive, high angular resolution X-ray imaging; \href{https://www.lem-observatory.org/}{\textit{LEM}} (a proposed US Probe mission; \citealt{LEM}) is designed to effectively map the thermodynamics and kinematics of the low-density CGM and WHIM using spectral imaging of soft X-ray line emission; \href{https://hubs-mission.cn/en/index.html}{\textit{HUBS}}, a Chinese mission with a X-ray microcalorimeter operating in the 0.1-2 keV band over a large field of view ($1 \times 1$ degree$^2$) with $<1$ arcmin resolution (expected launch: 2031) that will focus on the science of CGM, WHIM and filaments around galaxy clusters \citep[see, e.g.,][]{hubs25}.
 
The complementarity of SZ and X-ray measurements of the warm/hot content of cosmic large-scale structures has long represented a valuable asset for a cross-enhancement of the respective astrophysical information. The different dependence of these tracers on the physical properties of the ionized gas has been broadly exploited --- from, e.g. obtaining tighter constraints on the thermodynamics of the hot gas in distant clusters and cluster outskirts (e.g., \citealt{Ghirardini2019,Castagna2020,Ghirardini2021,Andreon2021,Ruppin2021,Lepore2023}) and large-scale filaments \citep[e.g.,][]{Planck2013VIII,Akamatsu2017,Hincks2022}, to studying local deviations from particle and thermal equilibrium \citep[e.g.,][]{Basu2016,DiMascolo2019b,Sayers2021}, deriving detailed morphological models of the three-dimensional distribution of ionized gas \citep[e.g.,][]{DeFilippis2005,Limousin2013,Umetsu2015,Sereno2018,Kim2024}, or obtaining measurements of the Hubble–Lemaître parameter independently of more standard probes (e.g., \citealt{Bonamente2006,Kozmanyan2019,Wan2021}).

The enhanced sensitivity, spatial resolution, and mapping speed of AtLAST for various flavors of the SZ effect, combined with the capabilities of next-generation X-ray facilities, will undoubtedly take these already existing synergies one leap further. On the other hand, the novel high spectral resolution imaging capabilities in the soft X-ray band (expected to become available in the next decade with, e.g., {\it NewAthena}) will give rise to new opportunities for complementary measurements with the SZ band. Namely, X-ray observations are primarily expected to map the line emission or absorption signals from {\it metals} in the diffuse, warm-hot gas permeating large-scale structure WHIM filaments or the low-mass haloes of individual L* galaxies (i.e., with a luminosity equal to the characteristic luminosity of the Schechter function; \citealt{Schechter1976}). The X-ray continuum emission from these targets will be swamped by the foreground continuum from our own Milky Way, and extremely difficult to probe (see, for instance, \citealt{Kraft2022}). The ideal path to obtaining a full picture of the physical properties of this diffuse gas component of the cosmic web, therefore, is to combine diagnostics about the metal content (from X-ray line intensities), metal dynamics (from X-ray line widths and shifts), temperature (from X-ray line ratios and relativistic SZ terms) with the gas pressure cleanly measured through the thermal SZ signal. We can then solve for the gas density (knowing the pressure and temperature), and the gas metallicity (knowing the metal content and gas density). Taking this one step even further, by detecting the kinetic SZ signal from the same gas, it will be possible to compare the velocities of metal-poor (primordial) gas from the kinetic SZ measurements which may be different than the velocities of metals probed from the X-ray lines. This will provide truly groundbreaking information about the circulation of gas and metals in and out of galaxies, by offering the opportunity to map, for instance, metal-rich outflows driven by feedback, and metal-poor inflows driven by accretion from the cosmic web, leading to a revolution in our understanding of galaxy evolution. 

\section{Summary and conclusions}
AtLAST will provide a transformational perspective on the SZ effect from the warm/hot gas in the Universe. The high angular resolution enabled by the 50-meter aperture, the extensive spectral coverage, and the extreme sensitivity swiftly achievable over wide areas of the (sub)millimeter sky will provide the unprecedented opportunity to measure the SZ signal over an instantaneous high dynamic range of spatial scales (from few arcsecond to degree scales) and with an enhanced sensitivity ($\lesssim 5\times 10^{-7}$ Compton $y_{\mathrm{t\textsc{sz}}}$).

Such a combination of technical advances will allow us to constrain simultaneously the thermal, kinematic, and relativistic contribution to the SZ effect for a vast number of individual systems, ultimately opening a novel perspective on the evolution and thermodynamics of cosmic structures. Such an unmatched capability will provide the means for exploring key astrophysical issues in the context of cluster and galaxy evolution. 
\begin{itemize}[leftmargin=*,topsep=2pt,itemsep=2pt,parsep=2pt]
\item By resolving the multi-faceted SZ footprint of galaxy clusters, low-mass groups, and protoclusters, it will be possible to trace the temporal evolution of their thermodynamic properties across (and beyond) the entire cluster era ($z\lesssim2$), over an unprecedented range in mass. The complementary information on the full spectrum of small-scale ICM perturbations that will be accessed thanks to AtLAST's superior resolution and sensitivity will thus allow us to build a complete picture of the many intertwined processes that make galaxy clusters deviate from the otherwise hydrostatic equilibrium and self-similar evolution. At the same time, we will be able to get a complete census of the cluster population, circumventing the inherent biases associated with current cluster selection strategies. Overall, such studies will allow AtLAST to be pivotal in firming the role of galaxy clusters as key cosmological probes.

\item The possibility offered by AtLAST of accessing the low-surface brightness regime will open an SZ window on the low-density warm/hot gas within the cosmic large-scale structure --- ranging from the characterization of the mostly unexplored properties of the assembling ICM seeds within protocluster overdensities to the barely bound outskirts of galaxy clusters. These represent the environments where the same process of virialization begins. As such, they are ideal for studying how deviations from thermalization, gas accretion, and strong dynamical processes impact the thermal history of galaxy clusters.

\item By tracing the imprint on the thermodynamics properties of circumgalactic medium surrounding galaxies and of the cluster cores, AtLAST will allow us to constrain the energetics and physical details of AGN feedback. This will provide the means for moving a fundamental step forward in our understanding of the crucial impact of AGN on the evolution of the warm/hot component of cosmic structures over a wide range of spatial scales and across cosmic history. 
\end{itemize}

To achieve these ambitious goals, it will be essential to satisfy the following technical requirements:
\begin{itemize}[leftmargin=*,topsep=2pt,itemsep=2pt,parsep=2pt]
    \item \textbf{Degree-scale field of view.} The superior angular resolution achievable thanks to the 50-meter aperture planned for AtLAST will need to be complemented by the capability of effectively recovering degree-level large scales. Such a requirement is motivated by the aim of mapping the SZ signal from astrophysical sources at low or intermediate redshift that are inherently extended on large scales (e.g., intercluster filaments) and with diffuse signals (e.g., protocluster overdensities). At the same time, we aim at performing a deep ($\sim10^{-7}$ Compton $y$) and wide-field ($>1000~\mathrm{deg^2}$) SZ survey, key for effectively probing a varied sample of SZ sources. In turn, our requirement consists of an instantaneous field of view covering $>1~\mathrm{deg^{2}}$. Clearly, combining wide-field capabilities with enhanced sensitivity will be highly demanding in terms of minimal detector counts. To reach the target sensitivities reported in Table~\ref{tab:freq_sens_beam}, we forecast that the focal plane array should be filled by $\gtrsim 50\,000$ detectors per spectral band.
    \item \textbf{Wide frequency coverage.} To perform a spectral inference of the multiple SZ components, along with their clean separation from foreground and background astrophysical contamination, it will be crucial to probe the spectral regime from $30~\mathrm{GHz}$ up to $905~\mathrm{GHz}$ with multi-band continuum observations. We specifically identify an overall set of nine spectral bands (centered at 42.0, 91.5, 151.0, 217.5, 288.5, 350.0, 403.0, 654.0, and 845.5~GHz), specifically selected to maximize the in-band sensitivity at fixed integration time. By testing this spectral configuration in the context of a mock spectral component separation, we demonstrated that such a choice allows for achieving a clean separation of multiple SZ components, as well as of the signal from dominant contamination sources.
    \item \textbf{Sub-percent beam accuracy.} An accurate calibration will be essential for reducing potential systematics in the small-amplitude fluctuations of the SZ signal associated with local pressure and velocity perturbations, or to relativistic distortions. As such, we require a sub-percent level control of the beam stability.
\end{itemize}

\section*{Data availability}
No data are associated with this article.

\section*{Software availability}
The calculations used to derive integration times for this paper were done using the AtLAST sensitivity calculator \citep{senscalc}, a deliverable of Horizon 2020 research project ``Towards AtLAST'', and available from \href{https://github.com/ukatc/AtLAST_sensitivity_calculator}{this link}.

\section*{Ethics and consent}
Ethical approval and consent were not required.
 
\section*{Competing Interests}
No competing interests were disclosed.

\section*{Grant information}
This project has received funding from the European Union’s Horizon 2020 research and innovation programme under grant agreement No.\ 951815 (AtLAST).
L.D.M.\ is supported by the ERC-StG ``ClustersXCosmo'' grant agreement 716762. L.D.M.\ further acknowledges financial contribution from the agreement ASI-INAF n.2017-14-H.0. This work has been supported by the French government, through the UCA\textsuperscript{J.E.D.I.} Investments in the Future project managed by the National Research Agency (ANR) with the reference number ANR-15-IDEX-01. 

Y.P.\ is supported by a Rutherford Discovery Fellowship (``Realising the potential of galaxy clusters as cosmological probes'') and Marsden Fast Start grant (``Turbulence in the Intracluster Medium: toward the robust extraction of physical parameters'').

S.A.\ acknowledges INAF grant  
``Characterizing the newly  discovered clusters of low surface  brightness'' and PRIN-MIUR 20228B938N grant
``Mass and selection biases of galaxy clusters: a multi-probe approach''. 

S.E.\ acknowledges the financial contribution from the contracts
Prin-MUR 2022 supported by Next Generation EU (n.20227RNLY3 {\it The concordance cosmological model: stress-tests with galaxy clusters}),
ASI-INAF Athena 2019-27-HH.0, ``Attivit\`a di Studio per la comunit\`a scientifica di Astrofisica delle Alte Energie e Fisica Astroparticellare'' (Accordo Attuativo ASI-INAF n.\ 2017-14-H.0),
and from the European Union’s Horizon 2020 Programme under the AHEAD2020 project (grant agreement n.\ 871158).

M.L.\ acknowledges support from the European Union’s Horizon Europe research and innovation programme under the Marie Sk\l odowska-Curie grant agreement No 101107795.

D.N.\ acknowledges funding from the Deutsche Forschungsgemeinschaft (DFG) through an Emmy Noether Research Group (grant number NE 2441/1-1).

T.Mo. acknowledges the support of L.\ Page.

S.W. acknowledges support by the Research Council of Norway through the EMISSA project (project number 286853) and the Centres of Excellence scheme, project number 262622 (``Rosseland Centre for Solar Physics'').

\section*{Acknowledgements}
We thank William Coulton for useful insights on the CMB systematics and recovery of larger scales, Matthieu Béthermin for the support with the analysis and processing of the SIDES simulations, Rémi Adam and Charles Romero for the details about the transfer function processing for NIKA/NIKA2 and MUSTANG/MUSTANG2, and Neelima Sehgal for the details on the planned CMB-HD spectral setup. We further thank Adam D.\ Hincks and the ACT collaboration for sharing the \textit{Planck} and ACT SZ data for Abell 399--401. 

\bibliographystyle{apj_mod}
\setlength{\parskip}{0pt}
\setlength{\bibsep}{0pt}
\bibliography{atlast_SZ_science}


\end{document}